\documentstyle[epsfig]{mn}
 
\begin{document} 
 
\newcommand{\sm}{\mbox{M}_{\sun}} 
\newcommand{\den}{$\mbox{M}_{\sun}$pc$^{-3}$} 
\newcommand{\kms}{km~s$^{-1}$} 
\newcommand{\tst}{\textstyle} 
\newcommand{\be}{\begin{equation}} 
\newcommand{\ee}{\end{equation}} 

\newcommand{\nsp}{N_{\rm pix}}
\newcommand{\nbl}{N_{\rm bl}}
\newcommand{\nsrc}{N_{\rm s}}
\newcommand{\ngal}{N_{\rm gal}}
\newcommand{\nsky}{N_{\rm sky}}
\newcommand{\sigi}{\sigma_i}
\newcommand{\fsee}{f_{\rm see}}
\newcommand{\asp}{A_{\rm pix}}
\newcommand{\aspt}{A_{\rm pix,T}}
\newcommand{\amin}{A_{\rm min}}
\newcommand{\amax}{A_{\rm max}}
\newcommand{\at}{A_{\rm T}}
\newcommand{\umax}{u_{\rm max}}
\newcommand{\ut}{u_{\rm T}}
\newcommand{\lsrc}{L_{\rm s}}
\newcommand{\texp}{T_{\rm exp}}
\newcommand{\te}{t_{\rm e}}
\newcommand{\tfw}{t_{\rm FWHM}}
\newcommand{\gp}{\Gamma_{\rm p}}
\newcommand{\gpo}{\Gamma_{\rm p}^{\rm obs}}
\newcommand{\gpoj}{\Gamma_{{\rm p},j}^{\rm obs}}
\newcommand{\gpj}{\Gamma_{{\rm p},j}}
\newcommand{\gc}{\Gamma_{\rm c}}
\newcommand{\dd}{D_{\rm l}}
\newcommand{\ds}{D_{\rm s}}
\newcommand{\vt}{V_{\rm t}}
\newcommand{\re}{R_{\rm e}}
\newcommand{\bvl}{\mbox{\boldmath $V_{\rm l}$}}
\newcommand{\bvs}{\mbox{\boldmath $V_{\rm s}$}}
\newcommand{\rhoh}{\rho_{\rm h}}
\newcommand{\rhod}{\rho_{\rm d}}
\newcommand{\rhob}{\rho_{\rm b}}
\newcommand{\rhol}{\rho_{\rm l}}
\newcommand{\rhos}{\rho_{\rm s}}
\newcommand{\lbump}{L_{\rm bump}}
\newcommand{\massb}{M_{\rm b}}
\newcommand{\rmax}{R_{\rm max}}
\newcommand{\mh}{m_{\rm h}}
\newcommand{\ms}{m_{\rm s}}
\newcommand{\mlo}{m_{\rm l}}
\newcommand{\mup}{m_{\rm u}}
\newcommand{\mtl}{M/L_B}

\title[Theory of pixel lensing towards M31 I]{Theory of pixel lensing
towards M31 I: the density contribution and mass of MACHOs}

\author[E.~Kerins et al.]{E.~Kerins$^1$, B.J.~Carr$^2$, N.W.~Evans$^1$,
P.~Hewett$^3$, E.~Lastennet$^2$,
\newauthor 
Y.~Le~Du$^4$, A.-L.~Melchior$^{2,5}$, S.J.~Smartt$^3$ and D.~Valls-Gabaud$^6$
\newauthor
(The POINT--AGAPE Collaboration)\\
$^1$Theoretical Physics, 1 Keble Road, Oxford OX1 3NP, UK\\
$^2$Astronomy Unit, School of Mathematical Sciences, Mile End Road,
    London E1 4NS, UK\\
$^3$Institute of Astronomy, Madingley Road, Cambridge CB3 0HA, UK\\
$^4$Laboratoire de Physique Corpusculaire et Cosmologie, Coll\`ege de France,
    11 Place Marcelin Berthelot, F-75231 Paris, France\\
$^5$DEMIRM UMR~8540, Observatoire de Paris, 61 Avenue
    Denfert-Rochereau, F-75014 Paris, France\\
$^6$Laboratoire d'Astrophysique UMR~CNRS~5572, Observatoire
    Midi-Pyr\'en\'ees, 14 Avenue Edouard Belin, F-31400 Toulouse, France
}

\maketitle

\begin{abstract} 
POINT-AGAPE is an Anglo-French collaboration which is employing the
Isaac Newton Telescope (INT) to conduct a pixel-lensing survey towards
M31. Pixel lensing is a technique which permits the detection of
microlensing against unresolved stellar fields. The survey aims to
constrain the stellar population in M31 and the distribution and
nature of massive compact halo objects (MACHOs) in both M31 and the
Galaxy.

In this paper we investigate what we can learn from pixel-lensing
observables about the MACHO mass and fractional contribution in M31
and the Galaxy for the case of spherically-symmetric near-isothermal
haloes. We employ detailed pixel-lensing simulations which include
many of the factors which affect the observables, such as non-uniform
sampling and signal-to-noise ratio degradation due to changing
observing conditions.  For a maximum MACHO halo we predict an event
rate in $V$ of up to 100 per season for M31 and 40 per season for the
Galaxy. However, the Einstein radius crossing time is generally not
measurable and the observed full-width half-maximum duration provides
only a weak tracer of lens mass. Nonetheless, we find that the
near-far asymmetry in the spatial distribution of M31 MACHOs provides
significant information on their mass and density contribution. We
present a likelihood estimator for measuring the fractional
contribution and mass of both M31 and Galaxy MACHOs which permits an
unbiased determination to be made of MACHO parameters, even from
data-sets strongly contaminated by variable stars. If M31 does not
have a significant population of MACHOs in the mass range $10^{-3}~\sm
- 1~\sm$ strong limits will result from the first season of INT
observations. Simulations based on currently favoured density and mass
values indicate that, after three seasons, the M31 MACHO parameters
should be constrained to within a factor four uncertainty in halo
fraction and an order of magnitude uncertainty in mass ($90\%$
confidence). Interesting constraints on Galaxy MACHOs may also be
possible. For a campaign lasting ten years, comparable to the lifetime
of current LMC surveys, reliable estimates of MACHO parameters in both
galaxies should be possible.
\end{abstract}

\begin{keywords}
dark matter --- galaxies: haloes --- galaxies: individual
(M31) --- Galaxy: halo --- gravitational lensing.
\end{keywords}

\section{Introduction} \label{s1}

\subsection{Conventional microlensing: landmarks and limitations}

The detection of the gravitational microlensing effect due to compact
objects in the Galaxy is undoubtedly one of the great success stories
in astrophysics over the past decade. Surveys have discovered
around 20 candidates towards the Magellanic clouds and several hundred
towards the Galactic Bulge \cite{uda94,alc97,ala97,lass99,alc00}. 
Amongst these candidates a number of exotic lensing phenomena have been
catalogued, such as parallax effects, binary lensing (including
spectacular examples of caustic-crossing events), and finite
source-size effects. These discoveries are facilitated by coordinated
follow-up campaigns such as PLANET \cite{alb98} and MPS \cite{rhie99}
which act on microlensing alerts broadcast by the survey
teams. The absence of certain microlensing signals has also yielded a
clearer insight into the nature of halo dark matter. The null detection
of short duration events towards the Large Magellanic Cloud (LMC) by
the EROS and MACHO surveys indicates that, for a range of plausible
halo models, massive compact halo objects (MACHOs) within the mass
interval $10^{-7} - 10^{-3}~\sm$ provide less than a quarter of the
dark matter \cite{alc98}. This is an important result when set against
the current insensitivity of other techniques to this mass range.

Despite these successes a number of unsolved problems remain. The
optical depth measured towards the Galactic Bulge is at least a factor
two larger than can be accommodated by theoretical models
(e.g. Bissantz et al. 1997; Sevenster et al. 1999). Towards the LMC
the rate of detected events is consistent with the discovery of a
significant fraction of the halo dark matter. However, the implied
lens mass range ($0.1 - 1~\sm$) is not easily reconciled with existing
constraints on baryonic dark matter candidates \cite{carr94}, though
the MACHOs need not necessarily be baryonic. Furthermore, the
discovery of two possible binary caustic-crossing events towards the
LMC and the Small Magellanic Cloud (SMC) has thrown into question the
very existence of MACHOs.  Their caustic-crossing timescales, which
provide an indicator of their line-of-sight position, seem to exclude
either as being of halo origin, a statistically unlikely occurrence if
the halo comprises a significant MACHO component \cite{ker99}.  As a
result, there is a growing body of opinion that all events observed so
far towards the LMC and SMC may reside in the clouds
themselves. However, this explanation is itself problematic because it
requires that the clouds must either have a higher MACHO fraction than
the Galaxy or comprise substantial but diffuse stellar components not
in hydrodynamical equilibrium (Evans \& Kerins 2000, and references
therein).

These problems highlight two principal constraints on the ability of
conventional microlensing experiments to determine the nature and
distribution of MACHOs in the halo. The first limitation is their
inefficiency in differentiating between lensing by MACHOs and
self-lensing by the source population, since for most events one
observes only a duration and a position on the sky. These observables
are only weakly correlated with the location of the events along the
line of sight. The second constraint is the limited number of suitable
lines of sight through the halo. Conventional microlensing surveys
require rich yet resolved stellar fields and are thus limited to just
two lines of sight, the LMC and SMC, with which to probe MACHOs. The
line of sight to the Galactic Bulge is dominated by bulge and disc
lensing. The paucity of halo lines of sight, together with the rather
weak dynamical and kinematical constraints on Galactic halo structure,
also diminishes the prospect of being able to decouple information on
the Galactic distribution function and MACHO mass function.

\subsection{Beyond the Galaxy: a new target, a new technique}

\begin{figure}
\begin{center}
\epsfig{file=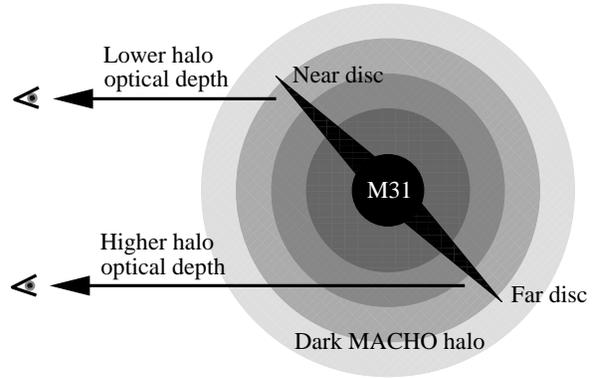,width=5cm,angle=270}
\end{center}
\caption{The principle of near-far asymmetry. The optical depth
through the halo towards the far disc is larger than towards the near
disc owing to the tilt of the disc confined within the spheroidal
distribution of MACHOs. The distribution of Galaxy MACHOs, disc
self-lensing events and variable stars does not exhibit asymmetry.}
\label{f1}
\end{figure}

The possibility of detecting MACHOs in an external galaxy,
specifically M31, was initially explored by Crotts (1992) and by
Baillon et al. (1993). Crotts (1992) pointed out that the high
inclination of the disc of M31 would result in an asymmetry in the
observed rate of microlensing if the disc is surrounded by a MACHO halo, as
illustrated in Figure~\ref{f1}. The fact that the M31 MACHO
microlensing rate should be lower towards the near side of the disc
than the far side, which lies behind a larger halo column
density, means that the presence of MACHOs in M31 can be established
unambiguously. In particular, neither variable stars nor stellar
self-lensing events in the disc of M31 should exhibit near-far
asymmetry. Additionally, the external vantage point serves to reduce
systematic model uncertainties in two ways. Firstly, it permits a
more accurate determination of the rotation curve and surface
brightness profile than is possible for the Galaxy, which reduces
the prior parameter space of viable galactic models. Secondly, it
provides many independent lines of sight through the halo of M31,
allowing the MACHO distribution across the face of the disc to be
mapped and thus the halo distribution function to be constrained more
or less directly.

As pointed out by Baillon et al. (1993), another appeal of directing
observations towards more distant large galaxies like M31 is the
increase in the number of potential source stars, more than a factor
of one thousand over the number available in the LMC and SMC, and all
confined to within a few square degrees.  However, this also presents
a fundamental problem in that the source stars are resolved only
whilst they are lensed (and even then only if the magnification is
sufficiently large). The presence of many stars per detector pixel
means it is often impossible to identify which is being
lensed. Furthermore, the flux contribution of the unlensed stars
dilutes the observed flux variation due to microlensing. Nonetheless,
Baillon et al. (1993) determined from numerical simulations that the
number of observable events, due to either the lensing of bright stars
or high magnification events, is expected to be large. As a result of
these studies, the Andromeda Galaxy Amplified Pixel Experiment (AGAPE)
and another group, Columbia-VATT, commenced observing programs towards
M31 \cite{ans97,cro97}.

One of the biggest technical difficulties facing surveys which look
for variable sources against unresolved stellar fields is how to
distinguish between flux variations due to changing observing
conditions and intrinsic variations due to microlensing or stellar
variability. For example, changes in seeing induce variations in
the detected flux within a pixel. One must also deal with the
consequences of positional misalignment between exposures, spatial and
temporal variations in the point spread function (PSF) and
photometric variations due to atmospheric transparency and variable 
sky background.

AGAPE has employed the Pixel Method to cope with the changing
observing conditions \cite{ans97}.  AGAPE thoroughly tested
this technique with a three-year campaign using the 2m Bernard Lyot
telescope at Pic du Midi from 1994 to 1996 \cite{ans97,ans99,ledu00}. Six
fields covering about 100 arcmin$^2$ centred on the bulge of M31 were
monitored. Whilst the field of view was insufficient to conclude much
about the nature of MACHOs, 19 candidate events were detected, though
it is still premature to rule out many of them being intrinsically
variable sources, such as Miras or novae. One event, AGAPE~Z1, appears
to be a convincing lensing candidate as its flux increase and colour
are inconsistent with that of a Mira or nova \cite{ans99}. A longer
baseline is needed to determine how many of the other candidates are
due to microlensing.

A major observing programme began on the 2.5m Isaac Newton Telescope
(INT) in La~Palma in the Autumn of 1999, with a run of one hour per
night for almost sixty nights over six months. The POINT-AGAPE
collaboration is a joint venture between UK-based astronomers and
AGAPE (where POINT is an acronym for ``Pixel-lensing Observations with
INT'').  We are exploiting the 0.3~deg$^2$ field of view of the INT
Wide-field Camera (WFC) to map the distribution of microlensing
events across a large region of the M31 disc.  Our initial
observations of M31 with the INT employed a $V$ filter and the
simulations reported here have been undertaken with parameters
appropriate to V-band observations. The strategy
employed for the actual M31 monitoring campaign involves observations
in three bands, $g$, $r$, and $i$ [very similar to the bands employed
by SLOAN \cite{fuk96}]. The multi-colour observations will improve our
ability to discriminate against variable stars and the $gri$-filter
plus CCD combination offers a significant improvement in sensitivity
(the $g$-band zero-point is some 0.4 magnitudes fainter than that for
$V$). The simulation parameters are thus somewhat conservative in this
regard. The programme is being conducted in consort with the
Microlensing Exploration of the Galaxy and Andromeda (MEGA) survey
\cite{cro99}, the successor program to Columbia-VATT. Whilst
POINT-AGAPE and MEGA are sharing the data, different techniques are
being employed to search for microlensing events. Henceforth we
use the term {\em pixel lensing}\/ \cite{gou96} to describe
microlensing against unresolved stellar fields, regardless of the
detection technique.

Whilst the technical viability of pixel lensing is now clearly
established, a number of important theoretical issues are still
outstanding. The principal concern is that the main observable in
classical microlensing, the Einstein crossing time, is generally not
accessible in pixel lensing. The Einstein crossing time is directly
related to the lens mass, its transverse velocity and the
observer--lens--source geometry. In pixel lensing the observed
timescale depends upon additional factors, such as the local surface
brightness and the source luminosity and magnification, so the
dependence on lens parameters is much weaker than for classical
microlensing.

The first detailed study of pixel lensing was undertaken by Gould
(1996). He defined two regimes: a semi-classical regime in
which the source star dominates the pixel flux and the observable
timescale provides a fair tracer of the Einstein crossing time; and
the ``spike'' regime where only high-magnification events are
identified, and the timescales are only weakly correlated with the
underlying Einstein crossing duration. Remarkably, Gould showed that,
despite the loss of timescale information, in the spike regime one can
still measure the microlensing optical depth. Using Gould's formalism,
Han (1996) provided the first pixel event rate estimates
for the M31 line of sight.  However, Gould's formalism assumes a fixed
sampling rate and unchanging observing conditions. As such it is of
limited applicability to a ground-based observing program. Gondolo
(1999) has proposed an optical depth estimator based on the
observed pixel event timescale. Whilst this estimator can be readily
employed by a ground-based campaign, it is somewhat sensitive to
the shape of the source luminosity function and is valid only to the
extent that this can be taken to be the same for
all source components.  More recently, Baltz \& Silk (1999)
derived expressions for the pixel rate and timescale distribution in
terms of the observable timescale, rather than the Einstein crossing
time. Again, their study assumes constant sampling and observing
conditions, as would be the case for space-borne programmes.

Whilst these studies provide a solid foundation for predictions of
pixel-lensing quantities (i.e. timescales, rates and optical depth),
none of them address to what extent one can constrain galactic and
lens parameters, in particular the MACHO mass, from pixel lens
observables.  Gyuk \& Crotts (2000) have shown that a reliable measure
of the optical depth from pixel lensing can be used to probe the core
radius and flattening of the M31 MACHO halo. 

In this paper we quantitatively assess the degree to which the
POINT-AGAPE campaign directed towards M31 will constrain the
fractional contribution and mass of the MACHOs.  Since the answer
inevitably depends upon the assumed galactic distribution function, we
focus attention here on the simple case of spherically-symmetric
near-isothermal halo models. The line of sight towards M31 is
sensitive to two MACHO populations, our own and that
in M31 itself, so we investigate the extent to which they can be
distinguished and probed independently. We also model the expected
background due to variable stars and lenses residing in the disc
and bulge of M31.

The plan of the paper is as follows. In Section~\ref{s2} we summarize
the basic principles of pixel lensing, with emphasis on the
differences between pixel lensing and classical microlensing. We
describe our Monte-Carlo pixel-lensing simulations in
Section~\ref{s3}, including our event selection criteria and the
incorporation of realistic sampling and observing conditions. In
Section~\ref{s4} we construct a reference model for the lens and
source populations in the halo of the Galaxy and the halo, disc and
bulge of M31, seeking consistency with the observed M31 rotation curve
and surface brightness profiles. In Section~\ref{s5} we present
predictions for the POINT-AGAPE survey based on our simulations. In
Section~\ref{s6} we use the simulations to generate artificial
data-sets and we investigate to what extent the MACHO mass and
fractional contribution in the two galaxies can be recovered from the
data. The results are summarized and discussed in Section~\ref{s7}.

\section{Principles of pixel lensing} \label{s2}

We review here some of the main aspects of pixel lensing
and its differences with classical microlensing. A more detailed
overview can be found in Gould (1996).

\subsection{Detecting pixel events}

Whilst in classical microlensing one monitors individual sources, in
pixel lensing the sources are resolved only whilst they are lensed. We
can therefore only monitor the flux in each detector element rather
than the flux from individual sources. If a star is magnified
sufficiently due to a lens passing close to its line of sight, then
the total flux in the detector element containing the source star (due
to the lensed star, other nearby unlensed stars and the sky
background) will rise significantly above the noise level and be
recorded as an event.

Before treating seeing variations the sequence of images must be
geometrically and photometrically aligned with respect to some
reference image, ${\cal R}$, as described in Ansari et al. (1997). The
variations remaining after alignment are primarily due to changes in
seeing and source flux, including microlensing events.  To minimize
the effects of seeing we define our base detector element to be a
superpixel: a square array of pixels.  A superpixel is defined for
each pixel, with that pixel lying at the centre, so that neighbouring
superpixels overlap with an offset of one pixel.  The optimal size for
the superpixel array is set by the ratio of the size of the seeing
disc on images obtained in poor seeing to the individual pixel
size. The INT Wide-field Camera (WFC) has a pixel scale corresponding
to $0\farcs 33$ on the sky, whilst poor seeing at La~Palma is $\sim
2\arcsec$. Adopting a very conservative value of $2\farcs 4$ for the
worst seeing leads to an optimized choice of $7 \times 7$ pixels for
the superpixel array. A larger array would overly dilute source
variations, whilst a smaller array would be overly sensitive to
changing observing conditions.

Whilst seeing variations are reduced by binning the photon count into
superpixels, this by itself is not enough to make them
negligible. Residual variations are minimized by the Pixel Method, in
which a simple, empirically-derived statistical correction is applied
to each image to match it to the characteristics of the reference
image ${\cal R}$. The Pixel Method is discussed in Ansari et al. (1997) and
described in detail by Le~Du (2000). The method strikes a good balance
between computational efficiency and optimal signal-to-noise ratio,
with the resulting noise level approaching the photon noise limit.

After alignment and seeing corrections the excess superpixel photon
count $\Delta \nsp$ on an image $i$ obtained at epoch $t_i$ due to an
ongoing microlensing event is
   \be
      \Delta \nsp(t_i) \equiv \nbl [\asp(t_i) -1 ] = \fsee \nsrc [A(t_i)-1].
      \label{spix}
   \ee
Here $\nsrc$ and $\nbl$ are the source and baseline photon counts in
the absence of lensing, $A$ is the source magnification factor due to
lensing and $\fsee$ is the fraction of the seeing disc contained
within the superpixel. The baseline photon count, $\nbl = \ngal ({\cal
R}) + \nsky ({\cal R})$, is the sum of the local M31 surface brightness
(including $\nsrc$) and sky background contributions on the reference
image. Whilst the quantities $\nbl$ and $\fsee \nsrc (A-1)$ can be
determined independently, $\nsrc$ and $A$ cannot in
general be inferred separately. It is therefore convenient to define
$\asp$ as the superpixel count variation factor, which acts as the
observable analogue of $A$.

The superpixel noise on image $i$ is 
   \be
      \sigi  =  \max [ \sigma_{\rm T}(t_i), \alpha_i 
      \nsp (t_i) ^{1/2} ], \label{err}
   \ee
where
   \be
      \nsp (t_i) = \Delta \nsp (t_i) + \nsky (t_i) + \ngal
      \label{supf}
   \ee
refers to the superpixel photon count on image $i$ {\em prior}\/ to
correction and, similarly, $\nsky$ and $\ngal$ are the uncorrected sky
background and galaxy surface brightness contributions. The threshold
noise level $\sigma_{\rm T}$ is determined by the superpixel flux
stability, and the scaling factor $\alpha_i$ takes account of the fact
that the Pixel Method is not photon-noise limited. A preliminary
analysis of a sequence of INT WFC images taken in 1998 demonstrated a
flux stability level of $0.1-0.3\%$ \cite{mel99}. We therefore adopt a
conservative minimum noise level of $\sigma_{\rm T} = 2.5 \times
10^{-3} \nbl$ for our simulations. We also apply a constant scaling
factor $\alpha_i = 1.2$, which is a little larger than typical for the
AGAPE Pic du Midi data
\cite{ledu00}. In reality $\alpha_i$ varies slightly between images
though we neglect this variation in our simulations.

Note that $\ngal$ in equation~(\ref{supf}) is constant, despite the
changing observing conditions. Though some variable fraction of the
local patch of surface brightness is dispersed over neighbouring
superpixels, the same amount of surface brightness leaks into the
superpixel from neighbouring patches, so there is no net
variation. The variation in $\nsky$ results from changing moonlight
and atmospheric transparency.

We regard a signal as being statistically significant if it occurs at
a level $3 \, \sigi$ above the baseline count $\nbl$.  Our estimate of
$\nbl$ must be obtained from a sequence of images and operationally is
defined to be the minimum of a sliding average of superpixel photon
counts over ten consecutive epochs. In order for a signal to be
detected on image $i$ we therefore require a superpixel count
variation factor $\asp (t_i) \ge 1+ 3 \, \sigi/\nbl$.  From
equation~(\ref{spix}), a microlensed source satisfies this inequality
provided that it is magnified by a factor exceeding
   \be
      \amin(t_i) = 1+ \frac{3 \, \sigi}{\fsee \nsrc}. \label{ampt}
   \ee
A special case of equation~(\ref{ampt}) occurs when $\sigi = \sigma_{\rm
T}$, giving a threshold magnification of
   \be
      \at = 1+ 0.0075 \frac{\nbl}{\fsee \nsrc}. \label{aspt}
   \ee
The extent to which residual temporal variations in $\fsee$ and $\nbl$
remain after image processing determines the factor by which $\sigi$
exceeds the photon noise limit, so this excess noise
is explicitly accounted for in equation~(\ref{ampt}).

Equation~(\ref{ampt}) illustrates some important characteristics of
pixel lensing. Firstly, pixel lensing does not depend directly on the
local surface brightness or sky background, but it does depend on
their contribution to the noise $\sigi$. Secondly, if the exposure
time $\texp$ is short, or the source star constitutes only a small
fraction of the superpixel flux, so that $\nsrc \ll \sigi$, only rare
high-magnification events are detected. The relationship between lens
magnification and lens--source impact distance (measured in the lens
plane) is as for the classical case:
   \be
      A = \frac{u^2 + 2}{u\sqrt{u^2 + 4}} \label{ampu}
   \ee
where $u$ is the impact distance in units of the Einstein radius. The
maximum value for the impact distance can be obtained by inverting
equation~(\ref{ampu}) for $A = \amin$:
   \be
      \umax  =  2^{1/2} \left[ \frac{\amin}{\sqrt{\amin^2 -1}} -1
      \right]^{1/2} \simeq  \amin ^{-1} \quad (\amin \ga 10).
      \label{imp}
   \ee
For pixel lensing in M31 we are often in the regime where $\nsrc \ll
\sigi$ because the source flux is much less than that of the galaxy
and background, so it is not unusual to require $\amin \ga 10$. In this
case equations~(\ref{ampt}) and (\ref{imp}) imply
   \be
      \umax \simeq \frac{\fsee \nsrc}{3 \, \sigi } < \frac{\fsee
        \nsrc}{3 \nsp^{1/2} } \quad \quad (\amin \ga 10), \label{ueq}
   \ee
Since $\umax \ll 1$ [typically $\umax \sim {\cal O}(10^{-2} -
10^{-3})$] only a small fraction of classical ($u \leq 1$)
microlensing events are detectable.

The dependence of $\umax$ on $\nsrc$ means that the pixel event rate
depends on the source luminosity function $\phi(M)$, the number
density of sources in the absolute magnitude interval $(M,M+dM)$. We
can compute a theoretical upper limit, $\gp$, for the pixel-lensing
rate at sky coordinate $(x,y)$ by taking $\amin = \at$ so that $\umax
 = u(\at) = \ut$. In this case
   \be
      \gp (x,y) = \langle \ut(x,y) \rangle_{\phi} \gc (x,y),
      \label{prate}
   \ee
where $x$ and $y$ are Cartesian coordinates centred on M31 and aligned
respectively along the major and minor axes of the projected light
profile. We define $y$ to be positive towards the near side of the
disc. The quantity $\gc$ is the classical ($u \leq 1$) event rate
integrated over lens and source populations \cite{grie91,kir94}, and
   \be
      \langle \ut (x,y) \rangle_{\phi} \equiv \frac{\int \ut(M,x,y)
        \phi(M) \, dM}{\int \phi(M) \, dM} \label{ulf}
   \ee
is the mean threshold impact parameter at $(x,y)$ averaged over
$\phi$.

Whilst useful in providing a rough order of magnitude estimate, $\gp$
cannot be compared directly with observations because it assumes
perfect sensitivity to all event durations and it also assumes that
observing conditions are unchanging. Since one usually has $\amin >
\at$, equation~(\ref{ulf}) also tends to overestimate the true mean
pixel-lensing cross-section. One can regard $\gp$, evaluated under the
best observing conditions, as providing a strict theoretical upper
limit to the observed event rate, in much the same way as $\gc$
provides an upper limit to the observed rate in classical lensing. In
Section~\ref{s3} we set about obtaining a more realistic estimate of
the observed pixel lensing rate.

\subsection{Degenerate and non-degenerate regimes} \label{snon-deg}

In classical microlensing the most important observable is the
Einstein radius crossing time, since this is directly related to the
position, motion and mass of the lens. Can we obtain similar
information from the duration of pixel events?

For a lens moving at constant velocity across the line of sight, $u$
evolves with time $t$ as in the classical case:
   \be
      u(t)^2 = u(t_0)^2 + \left( \frac{t - t_0}{\te} \right)^2,
      \label{utime}
   \ee
where $t_0$ is the epoch of minimum impact distance and $\te$ is the
Einstein radius crossing time. From equations~(\ref{ampu}) and
(\ref{utime}), $\te$ gives the timescale over which the source
magnification $A$ varies significantly. For large magnifications $u
\simeq A^{-1}$ from equation~(\ref{imp}), and inserting
equation~(\ref{utime}) into equation~(\ref{spix}) gives
   \be
      \Delta \nsp(t) \simeq  \frac{ \fsee \amax \nsrc }{ \sqrt{1 + \left(
      \frac{ {\tst t-t_0} }{ {\tst \te \amax^{-1} } } \right)^2} } \quad
      \quad [A(t) \ga 10], \label{lcurve}
   \ee
where $\amax \equiv A(t_0)$ is the maximum magnification. We infer
that in pixel lensing the timescale over which the signal varies
significantly is $\te \amax^{-1}$ rather than $\te$. This means that,
in the high-magnification regime, the pixel-lensing timescale bears
little relation to $\te$. We also see that the light-curve is
degenerate under transformations $\amax \rightarrow \alpha \amax$, $\nsrc
\rightarrow \nsrc/ \alpha$ and $\te \rightarrow \alpha \te$
\cite{woz97}. So neither $\te$, $\amax$ nor $\nsrc$ can be determined
independently. It may sometimes be possible to break this degeneracy
by looking at the wings of the light-curve \cite{bal99}, where
differences between the true magnification and its degenerate form can
become apparent. From equation~(\ref{ampu}), the difference between
the exact expression for $A(u)-1$ appearing in equation~(\ref{spix})
and its degenerate approximation, $u^{-1}$, is 
   \be
      \Delta (A-1) = \frac{u^2 + 2}{u\sqrt{u^2 + 4}} - 1 - \frac{1}{u}
      \simeq \frac{3u}{8} - 1 \quad \quad (u \la 1).
      \label{difa}
   \ee
To discriminate reliably (say at the $3 \, \sigma$ level) between the
degenerate and non-degenerate cases requires $\fsee \nsrc |\Delta
(A-1)| > 3\, \sigi$, so for the high-magnification regime we can
write the condition for non-degeneracy as
   \be
      \sigi \la \frac{\fsee \nsrc}{3}
      \quad \quad (u \ll 1). \label{uint}
   \ee
Equation~(\ref{uint}) demands that the superpixel noise be no greater
than the contribution of the unlensed source to the superpixel flux.
In general this will not be the case, so observations will not be able
to break the light-curve degeneracy and thus will not directly probe
the Einstein crossing time.

Since the underlying duration $\te$ is not generally measurable we use
the observed full-width half-maximum (FWHM) event duration:
   \be
      \tfw = 2 \sqrt2 \, \te \, \left[ \frac{a+2}{\sqrt{a^2+4a}} -
      \frac{a+1}{\sqrt{a^2+2a}} \right]^{1/2}, \label{tfw}
   \ee
where $a = \amax-1$. Since $\amax$ for detected events is typically
larger in regions of higher surface brightness, and for fainter stars,
$\tfw$ is correlated both with the disc surface brightness and the
source luminosity function. This means that it is less strongly
correlated than $\te$ with the lens mass and velocity and the lens and
source distances.

The observed duration, $\tfw$, does not afford us with as direct a
probe of lens parameters as $\te$. We are therefore forced to rely on
other observables, such as spatial distribution, in order to probe the
underlying MACHO properties. For M31 MACHOs one can test for near-far
asymmetry in the event rate \cite{cro92}. For Galaxy MACHOs there is
no comparable signature. Looking from the centre of the Galaxy towards
M31 the halo density distribution in the two galaxies is highly
symmetric about the observer--source midpoint. Since the microlensing
geometry is also symmetric about the midpoint the timescale
distributions for Galaxy and M31 MACHOs are similar for the same mass
function.  Since our displacement from the Galactic centre is only
8~kpc (small compared to the scale of the haloes and the Galaxy--M31
separation) this geometrical symmetry is largely preserved at our
location. However, the Galaxy MACHO distribution ought to be less
concentrated than that of stellar lenses. One might hope to see this
as an excess of events at faint isophotes which remains the same
towards both the near and far disc.  If MACHOs exist, the overall
pixel-lens distribution will be superposition of several lens
populations (Galaxy halo, M31 halo, disc and bulge) together with
variable stars which, at least in the short term, appear
indistinguishable from microlensing. The task of disentangling each is
therefore potentially tricky.

\section{Simulating pixel events} \label{s3}

A straightforward method for probing the lens populations is to
construct simulations of the expected distribution of events for a
particular telescope configuration, set of observing conditions and
selection criteria and then compare these predictions to
observations. To this end we have constructed a detailed simulation of
a realistic pixel-lensing experiment.

Our simulation works by first computing a theoretical upper limit to
the pixel rate for assumed M31 and Galaxy models. This estimate
provides the basis for generating trial pixel microlensing events for
which light-curves are constructed and selection criteria applied. The
precise details of our input galaxy models are discussed in
Section~\ref{s4}; in this section we lay down the general framework
for the simulation. For each generated trial event, a pixel
light-curve is constructed using a realistic distribution of observing
epochs interrupted by poor weather and scheduling constraints. The
effects of the sky background and seeing are explicitly taken into
account in computing flux realizations and errors for each
``observation''. The observing sequence is then examined to see
whether the event passes the detection criteria --- if it does, then
the trial counts as a detected event. The simulation is terminated
once $10^4$ events are detected or $10^6$ trials generated, whichever
is reached first. The fraction of trial events which are detected is
used to compute the observed pixel rate. The statistical error on the
rate determination is typically about $3\%$.

\subsection{Generating trial events} \label{s3.1}

As the starting point for our simulation we use the theoretical pixel
event rate as a function of position, $\gp (x,y)$, defined by
equation~(\ref{prate}). This quantity, evaluated for the best seeing
conditions, always provides an upper limit to the detection rate at a
given location and is therefore convenient to use to generate trial
events. We compute $\gpj$ over a grid of locations $(x,y)$ for each
combination $j$ of lens and source population. Near the centre of M31,
$j = 1 \ldots 8$ since there are two source populations (M31 disc and
bulge) and four lens populations (Galaxy halo, M31 halo, M31 disc and
M31 bulge). Beyond 8~kpc the M31 bulge is not in evidence, so $j = 1
\ldots 3$.

Given the grid of $\gpj (x,y)$, one can write the probability of
observing an event at location $(x,y)$ as
   \be
      P(x,y) \propto \Delta x \Delta y \sum_j S_j (x,y) \gpj(x,y),
      \label{pxy}
   \ee
where $S_j$ is the source surface density at $(x,y)$ for lens--source
configuration $j$, and $\Delta x$ and $\Delta y$ are the local $x$ and
$y$ grid spacings (required only for non-uniform grids). $P(x,y)$
therefore reflects the total event rate in a box of area $\Delta x
\Delta y$ centred on $(x,y)$. The box should be sufficiently small
that $S_j (x,y)$ and $\gpj (x,y)$ provide good estimates of the source
density and theoretical rate anywhere within it. Having fixed the
event location, $\gpj$ is then used to select the lens and source
components from the probability distribution
   \be
      P(j) = \frac{S_j (x,y) \gpj (x,y)}{\sum_j S_j (x,y) \gpj (x,y)}.
      \label{pi}
   \ee

Once the event location and lens and source populations have been
decided, the next choice is the line-of-sight distances to the lens,
$\dd$, and source, $\ds$:
   \begin{eqnarray}
        P(\ds) & \propto & \rhos (\ds) \ds ^{3/2} \int_0^{\ds} P( \dd)
        \, d \dd \nonumber \\
        P(\dd) & \propto & \rhol (\dd) \sqrt{\dd ( \ds - \dd)},
        \label{posn}
   \end{eqnarray}
where $\rhol$ and $\rhos$ are respectively the lens and source mass
densities. These distributions reflect the dependency of the
microlensing rate $\gpj$ on $\ds$, integrated over all possible $\dd$,
and on $\dd$, for a given $\ds$. Next we require the lens mass $m$
and relative transverse speed $\vt$. The lens mass realization is
generated from the distribution
   \be
      P(m) \propto m^{1/2} \psi (m) \label{mfreal},
   \ee
since, in the absence of finite source-size effects, $\gp \propto \re
\psi \propto m^{1/2} \psi$, where $\psi$ is the lens mass function
(i.e. the number density of lenses per unit mass interval) and $\re$
is the Einstein radius. The transverse speed $\vt (\bvl,\bvs)$ is
drawn from the assumed velocity distributions $P_{\rm l}(\bvl)$ and
$P_{\rm s}(\bvs)$ (see section~\ref{s4}), with $\bvl$ and $\bvs$ the
lens and source three-dimensional velocity vectors. Since the
microlensing rate $\gp$ is proportional to $\vt P_{\rm l} P_{\rm s}$
rather than just $P_{\rm l} P_{\rm s}$, each of our realizations must
be weighted by $\vt$ in computing the final detection rate. Finally,
we also need to generate the source absolute magnitude $M$ (defined
for some photometric band). The dependency of $\gp$ on $M$ derives
from the luminosity function $\phi$ and the threshold impact parameter
$\ut$, so we have
   \be
      P(M) \propto \ut (M,x,y) \phi(M). \label{mreal}
   \ee

\subsection{Generating light-curves} \label{s3.2}

At this point we have only simulated events according to the
underlying distributions which govern $\gp$; we have yet to take into
account the distribution of observing epochs, variations in observing
conditions, or candidate selection criteria.

The observing season runs from the beginning of August to the end of
January, so we adopt the duration of an observing season to be $\Delta
T = 180$~days. We assume 60 scheduled observing epochs per season ---
approximately the number of nights awarded for our 1999/2000
season. To construct a realistic sequence of observing epochs we
assume that the WFC is mounted on the telescope and available for
two-week periods every four weeks and that, on average, $25\%$ of
scheduled observations are precluded by bad weather. Periods of poor
weather are superposed on our initial observing sequence to obtain a
final sequence which typically comprises 40--50 epochs per season. In
practice we expect to obtain observations on more epochs than this,
but for the purposes of these simulations we assume 40--50 as a
conservative lower limit. For example during the 1999/2000 season we
have had observations on 56 nights.

The epoch of maximum magnification $t_0$ and the minimum impact
parameter $u(t_0)$ are both chosen at random. $u(t_0)$ is selected
from the interval $[0,\ut]$, where the threshold impact parameter
$\ut$ is computed from equations~(\ref{aspt}) and (\ref{imp}) taking
$\amin = \at$. This is all that is required to generate
the underlying microlensing light-curve. 

To compute the pixel light-curve, we must also model the galaxy
surface brightness and sky background. The simulations presented here
are performed in the $V$ band and we use the radially-averaged surface
brightness profile in Table~VI of Walterbos
\& Kennicutt (1987) to estimate the contribution to the pixel flux of
the galaxy background at the event location. The assumed sky
background corresponding to a dark sky is listed in Table~\ref{t1},
along with other INT detector and site characteristics. The sky
background varies over lunar phase and we adopt a contribution to the
sky background from the full moon equivalent to $10^3$ tenth magnitude
stars per deg$^2$ (c.f. Krisciunas \& Schaefer 1991). The contribution
is modulated according to the lunar phase. The lunar contribution to
the sky background also depends upon whether the moon is above the
horizon and on its angular distance from M31. Our assumed value is
taken to be an average over the positional dependence, so the true
variation in the sky background will be somewhat larger than we
consider. We also simplify the computation of the seeing fraction
$\fsee$ by adopting a Gaussian PSF with a FWHM equal to the seeing of
the reference image. The position of the PSF maximum for the reference
image is selected at random within the central pixel of the superpixel
array.
\begin{table}
\begin{center}
\caption{Adopted characteristics of the INT observing site and
Wide-field Camera (WFC). The sky background is given in
mag~arcsec$^{-2}$ and the superpixel dimension is quoted in
pixels. The zero-point is given in terms of the apparent magnitude of
a source which results in a 1~photon~sec$^{-1}$ detection rate. All
magnitudes are for the $V$ band.  Our survey is now observing in $g$,
$r$ and $i$ filters. For comparison, the sky background and zero-point
in $g$ are 22.2 and 26.0, respectively.}
\label{t1}
\begin{tabular}{@{}lc}
Characteristic & INT WFC \\
\hline
Best seeing & $0 \farcs 8$ \\
Worst seeing & $2 \farcs 4$ \\
Reference image seeing & $1 \arcsec$ \\
Sky background & 21.9 \\
Scheduled epochs per season & 60 \\
Field dimensions & $32\arcmin \times 32\arcmin$ \\
Pixel field of view & $0\farcs 33$ \\
Superpixel dimensions & $7 \times 7$ \\
Zero-point & 25.6 \\
Exposure time per field & 760~secs \\
\end{tabular}
\end{center}
\end{table}

Using our computed values for $\fsee$, the INT
detector and site characteristics summarized in Table~\ref{t1}, and the
microlensing parameters generated for each event, we construct
superpixel light-curves via equation~(\ref{spix}). The error at each epoch
$i$ is given by equation~(\ref{err}). Poisson realizations for the
superpixel flux at each epoch are generated from $\nsp (t_i)$ and
$\sigi$.

\subsection{Selection criteria and the observed rate} \label{s3.3}

\begin{figure*}
\centering
\begin{minipage}{170mm}
\begin{center}
\epsfig{file=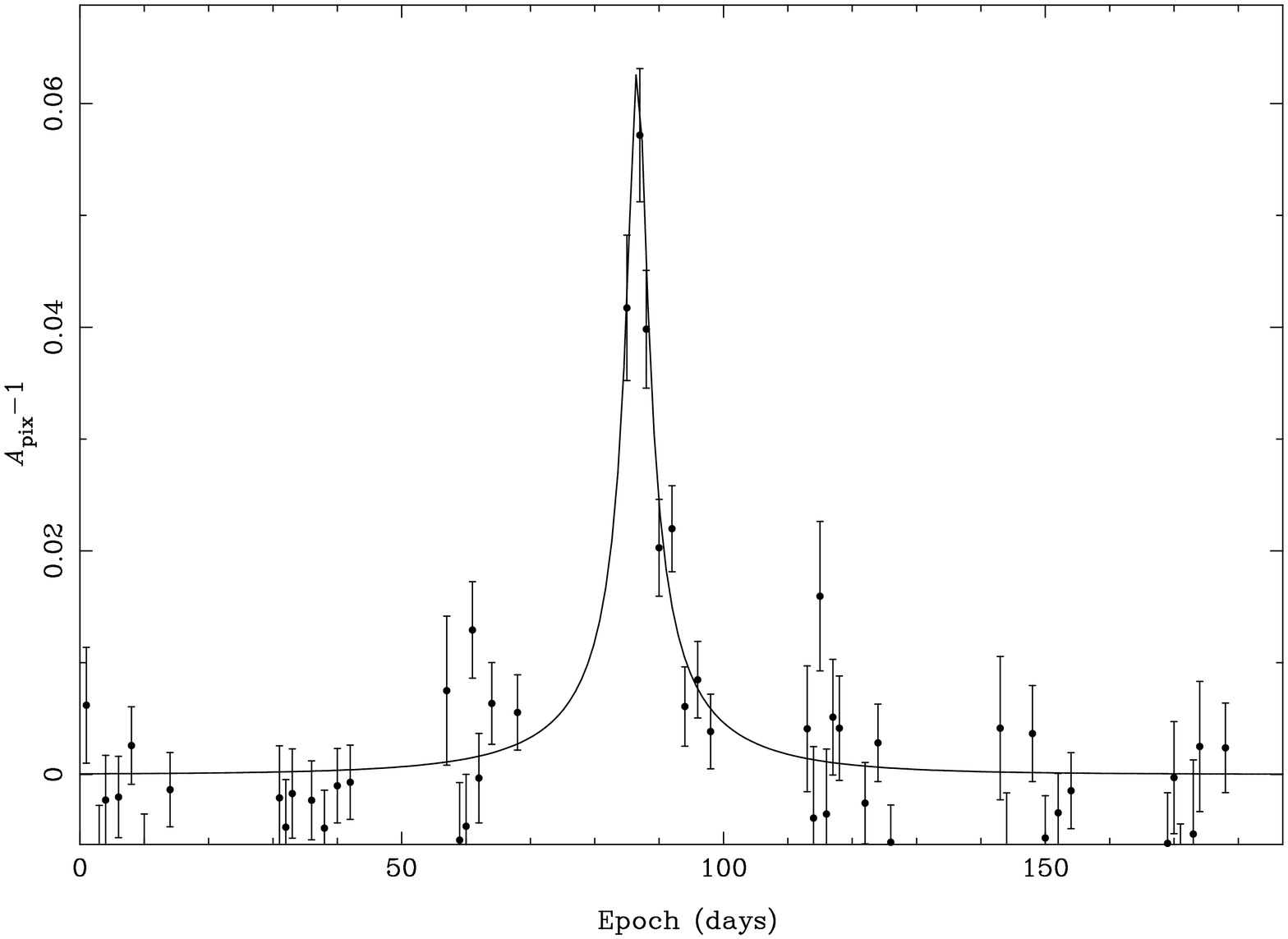,width=7cm,angle=270}
\epsfig{file=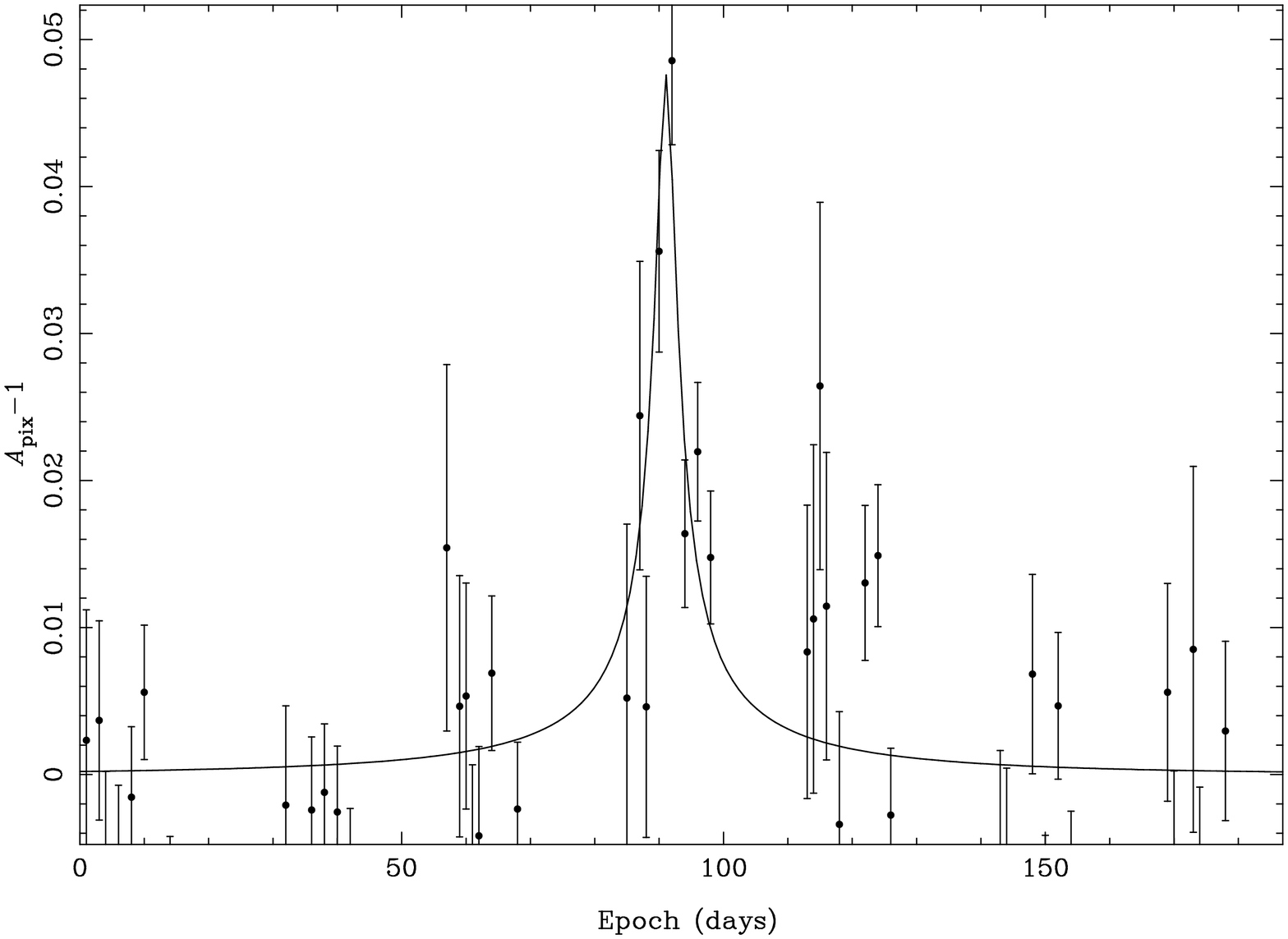,width=7cm,angle=270}
\epsfig{file=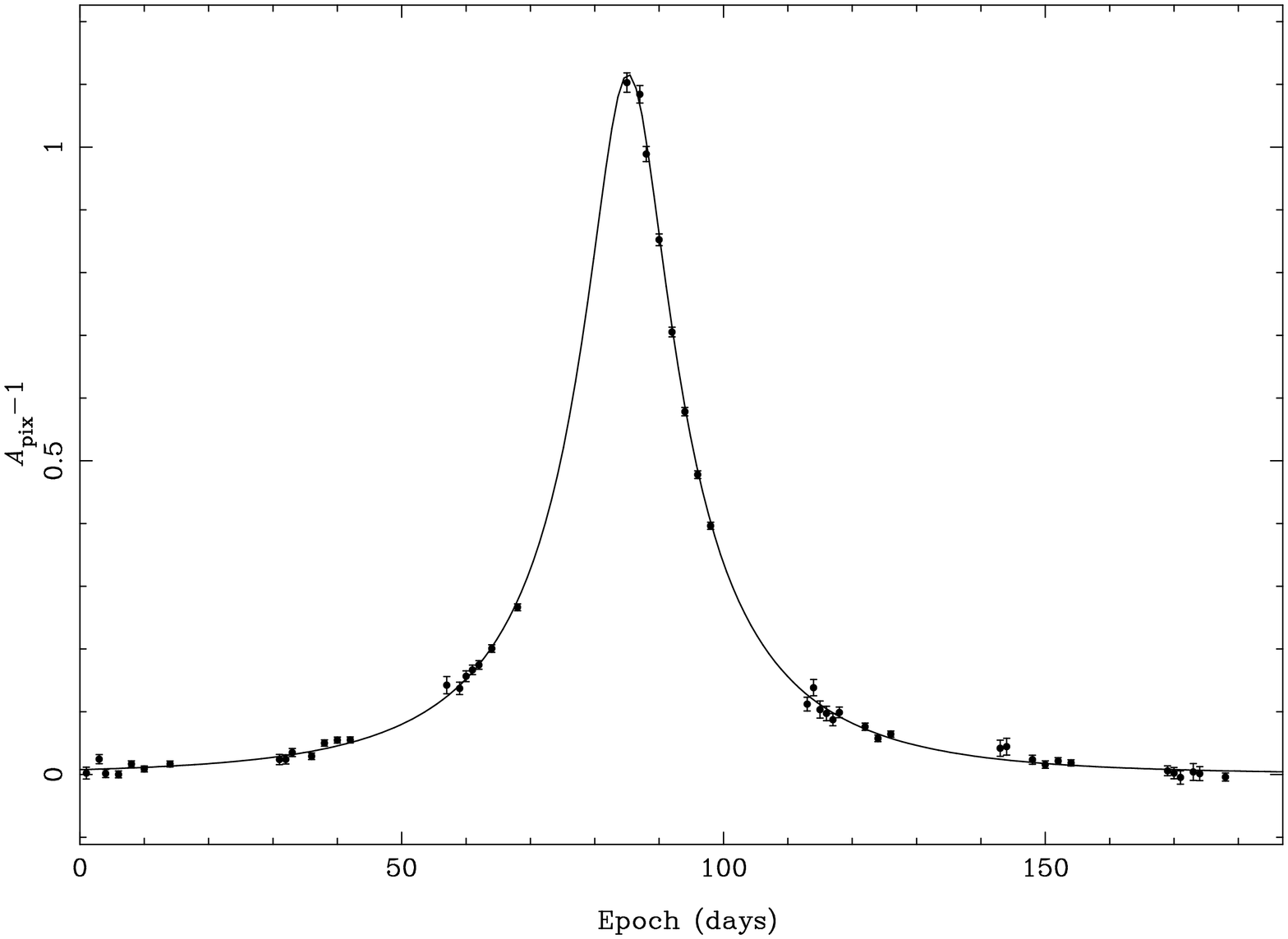,width=7cm,angle=270}
\end{center}
\caption{Simulated pixel-lensing light-curves. ({\it Top panel}\/) A
light-curve with a signal-to-noise ratio typical of many of the events;
({\it middle panel}\/) a low signal-to-noise ratio event; ({\it bottom
panel}\/) a high signal-to-noise ratio light-curve.}
\label{f3}
\end{minipage}
\end{figure*}
\begin{table*}
\centering
\begin{minipage}{170mm}
\caption{Parameters adopted for the density and velocity distributions
for components of the Galaxy and M31. The bulge model is adopted from
Kent (1989).}
\label{t2}
\begin{center}
\begin{tabular}{@{}lllll}
\null & \null & \null & Rotation & Velocity \\
Component & Mass normalization & Scale lengths & speed & dispersion \\
\hline
M31 bulge & $\massb = 4 \times 10^{10}~\sm$ & --- & 30~\kms & 100~\kms \\
\null & $\mtl$ = 9 & \null & \null & \null \\
M31 disc & $\rhod (0) = 0.2$~\den & $H = 0.3$~kpc, & 235~\kms & 30~\kms \\
\null & $\mtl$ = 4 & $h = 6.4$~kpc & \null & \null \\
M31 halo & $\rhoh (0) = 0.23$~\den & $a = 2$~kpc, & 0 & 166~\kms \\
\null & \null & $\rmax = 200$~kpc & \null & \null \\
Galaxy halo & $\rhoh (0) = 0.036$~\den & $a = 5$~kpc, & 0 & 156~\kms \\
\null & \null & $\rmax = 100$~kpc & \null & \null
\end{tabular}
\end{center}
\end{minipage}
\end{table*}
The adoption of selection criteria inevitably reduces the number of
detected events, but they are necessary to minimize the number of
contaminating non-microlensing signals. As in all microlensing
experiments the selection criteria must be based upon the quality of
the data and the characteristics of non-microlensing
variations. Ultimately the criteria must be derived from the data
themselves, so they are inevitably experiment-specific and evolve as
the experiment progresses. For our simulations we impose criteria
based loosely on the previous AGAPE pixel-lensing at Pic du Midi
\cite{ans97,ledu00}.

The principal criterion for the selection of microlensing events in
our simulation is that one and only one significant bump be identified
on the light-curve. The bump must comprise at least three consecutive
measurements lying at least $3 \sigma$ above the baseline superpixel
flux. Quantitatively, the significance of a bump is defined by its
likelihood
   \be
      \lbump = \prod_{i = j}^{i = j+n, n \geq 3} P(\Theta > \Theta_i |
      \Theta_i \geq 3), \label{bumplike}
   \ee
where $\Theta_i = [\nsp (t_i) - \nbl]/\sigi$ and $P(\Theta)$ is the
probability of observing a deviation at least as large as $\Theta$ by
chance. For a Gaussian error distribution, $P = \frac{1}{2}
\mbox{erfc} (\Theta / \sqrt{2} )$. Equation~(\ref{bumplike})
indicates that we evaluate $P(\Theta_i)$ only when $\Theta_i \geq
3$. For our simulations we demand that a candidate have one bump with
$- \ln \lbump > 100$ and no other bump with $- \ln \lbump > 20$. We
further demand that the epoch of maximum magnification $t_0$ lies
within an observing season; we reject candidates which attain their
maximum brightness between seasons, even if they last long enough for
the tails of the light-curve to be evident. This helps to ensure a
reliable estimate of the peak flux, and in turn the FWHM timescale
$\tfw$.

The bump criterion is both a signal-to-noise ratio condition and a
test for non-periodicity. It is crucial for distinguishing
microlensing events from periodic variables, though long-period
variables, such as Miras, may pass this test in the short term. In
addition to the bump test, one can also test the goodness of fit of
the light-curve to microlensing, which helps to distinguish microlensing
from typical novae light-curves. Though the presence of the background
means that pixel events will not in general be achromatic, the ratio
of the flux increase to baseline flux in different colours should
nonetheless be independent of time, so this provides another test for
microlensing. Colour information may also help to exclude some
long-period variables in the absence of a sufficient baseline of
observations. In Section~\ref{s6} we also exploit differences in
spatial distribution to separate statistically lensing events from
variable stars.

For real data-sets we would require more criteria in order to avoid
excessive contamination from variable stars.  For now we are
simulating only microlensing events, so we are assured of no
contamination in our selection. However, the cuts adopted above would
be responsible for many of the rejected candidates in a real
experiment, so the absence of further criteria should not lead to a
gross overestimate of the rate. In any case, we have been deliberately
conservative with our choices of sky background level, worst seeing
scale, the number of epochs per season and the pixel stability level
$\sigma_{\rm T}$. We therefore feel our predictions are more likely to
be underestimates of the actual detection rate.

The observed rate can be now readily computed from $\gp$, the number
of generated trials and the fraction of these which pass the detection
criteria. As mentioned in Section~\ref{s3.1}, the way in which
velocities are generated in the simulations means that the correct
rate is obtained by weighting each event by its transverse speed
$\vt$. Thus, the observed rate for lens component $j$ is
   \be
      \gpoj = \langle \gpj \rangle_{x,y} \frac{\sum_{l = 1}^{N_{\rm det}(j)}
      V_{{\rm t},l} }{ \sum_{k = 1}^{N_{\rm trial}(j)} V_{{\rm t},k} },
      \label{robs}
   \ee
where $\langle \gpj \rangle_{x,y}$ is the spatial average of $\gpj$
(summed over source populations), the lower summation is over all
$N_{\rm trial}$ trial events generated for lens component $j$ and the
upper summation is over the $N_{\rm det}$ detected events which pass
the selection criteria. The total number of events after $n$ observing
seasons is
   \be
      N =  n \, \Delta T \, 10^{0.4(\langle M
      \rangle - M_{\rm gal})} \sum_j \gpoj, \label{nexp}
   \ee
where $\langle M \rangle$ is the average absolute magnitude of the
sources (integrated over the luminosity function) and $M_{\rm gal}$ is
the absolute magnitude of M31 ($M_V = -21.2$).

\subsection{Simulated light-curves} \label{s5.1}

Three light-curves generated for a first-season simulation involving
$0.1~\sm$ MACHOs are shown in Figure~\ref{f3}. The galactic models
required for the simulation are discussed in Section~\ref{s4}. The
light-curves illustrate the range in signal-to-noise ratio. The
down-time for the WFC is evidenced by the way in which the epochs are
clumped into two-week periods. The variation in the size of the error
bars reflects the simulated variation in observing conditions.

Figure~\ref{f3}a shows an M31 halo lens magnifying a bulge star ($M_V
= -0.4$) and is a typical example. The underlying maximum
magnification for this event is $\amax = 18$, whilst the maximum
enhancement in superpixel flux is $\asp (t_0) = 1.06$, indicating that
the unlensed source is contributing less than $0.4\%$ of the
superpixel flux. For this event $\tfw = 5$~days and $\te = 28$~days.
Figure~\ref{f3}b, which illustrates a poor candidate with a low
signal-to-noise ratio, involves a Galaxy MACHO and $M_V = 1.8$ bulge
source contributing only $0.1\%$ of the superpixel flux ($\amax = 42$,
$\asp (t_0) = 1.05$). In this example $\tfw = 5$~days and $\te =
68$~days. Though there appears to be evidence of a second bump after
the main peak these points are all within $3\, \sigma$ of the baseline
and so do not count as a bump. Figure~\ref{f3}c shows a high
signal-to-noise ratio ``gold-plated'' event in which a very luminous
($M_V = -4$) disc source is lensed by an M31 MACHO ($\amax = 5$, $\asp
(t_0) = 2.1$) with an observed duration $\tfw = 19$~days and
underlying timescale $\te = 33$~days. Here the bright unlensed source
accounts for $27\%$ of the superpixel flux.

\section{Lens and source models} \label{s4}

\begin{figure*}
\centering
\begin{minipage}{170mm}
\begin{center}
\epsfig{file=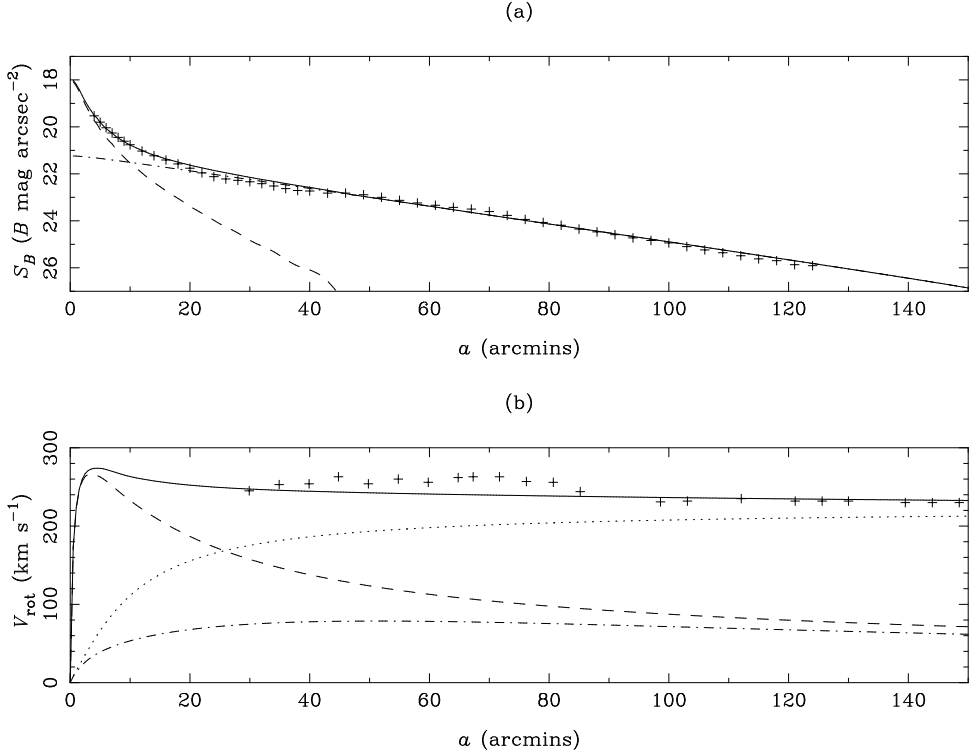,width=10cm,angle=270}
\end{center}
\caption{(a) The overall surface brightness profile (solid line) as a
function of semi-major axis $a$ for our M31 model produced by the
combined bulge (dashed line) and disc (dot-dashed line) light. The
crosses are radially-averaged measurements from Table~VI of Walterbos
\& Kennicutt (1987). (b) The overall rotation curve (solid line) for
the same M31 model summed over bulge (dashed line), disc (dot-dashed
line) and halo (dotted line) contributions. The crosses are from
Figure~2 of Kent (1989) and are based on emission line measurements.
For conversion to distance $1~\mbox{kpc} = 4.5$~arcmin.}
\label{f2}
\end{minipage}
\end{figure*}
In order to make quantitative estimates for pixel-lensing observables, we
must specify models for the principal Galaxy and M31 lens and
source components. For M31 the main populations are the bulge,
the disc and the dark MACHO halo. For the Galaxy only the MACHO
halo is important since the disc does not contribute
significantly. Our complete model therefore consists of these four
populations. Two populations, the M31 disc and bulge, also provide the
sources, so in total we have eight different lens--source
configurations. For each population we must specify distributions for
the density and velocity. Additionally, we must specify the lens mass
and a luminosity function for the source populations.  Throughout we
assume a disc inclination of $77\degr$ and a distance to M31 of
770~kpc, consistent with recent determinations (e.g. Stanek \&
Garnavich 1998).

Whilst the present paper is concerned only with quantities
relating to M31 and Galaxy MACHOs, we must nonetheless include other
significant lens components in our modeling in order to properly
characterize the complexity of extracting physical information from
observations. For the observations, unlike the simulations, we do not
know in which population a particular lens resides.

The haloes are modeled as simple near-isothermal spheres with cores, having
density profiles
   \be
      \rhoh = \left\{ \begin{array}{ll} 
        \rhoh (0) \frac{\tst a^2}{\tst a^2 + r^2} & (r \leq \rmax) \\
        0 & (r > \rmax)
      \end{array} \right., \label{halod}
   \ee
where $\rhoh (0)$ is the central density, $a$ is the core radius,
$\rmax$ is the cutoff radius and $r$ is the radial distance measured
from the centre of either M31 or the Galaxy. The assumed values for
$\rhoh (0)$, $a$ and $\rmax$ are given in Table~\ref{t2}. The halo
fraction determinations in Section~\ref{s6} are made with respect to
these density normalizations. In our model the M31 halo has about twice the
mass of the Galactic halo, though this mass ratio is controversial and
has been challenged recently by Evans \& Wilkinson (2000) who have
studied the kinematics of several satellite galaxies around M31.

The M31 disc is modeled by the sech-square law:
   \be
      \rhod = \rhod (0) \exp \left( - \frac{\sigma}{h} \right) {\rm
sech}^2
      \left( \frac{z}{H} \right), \label{discd}
   \ee
where $\sigma$ is the radial distance measured in the disc plane and
$z$ is the height above the plane. The normalization $\rhod (0)$,
scale-height $H$ and scale-length $h$ are given in Table~\ref{t2}.

The bulge distribution is based on the work of Kent (1989). Kent
models the bulge as a set of concentric oblate-spheroidal shells with
axis ratios which vary as a function of semi-major axis. We use the
tabulated spatial luminosity density values in Table~1 of Kent (1989)
and normalize the bulge mass under the assumption that the light
traces the mass (constant bulge mass-to-light ratio). The mass
normalization $\massb$ is listed in Table~\ref{t2}. The assumption of
axisymmetry may be over-simplistic since the misalignment between the
disc and bulge position angles probably implies a triaxial structure
for the bulge. However, we are only indirectly concerned with bulge
lensing in so much as it contaminates halo lensing statistics, so
deviations from axisymmetry are not crucial.

The rotation curve and surface brightness profile for the adopted M31
components are shown in Figure~\ref{f2}. In constructing the surface
brightness profile, we have assumed $B$-band mass-to-light ratios $\mtl
= 4$ for the disc and $\mtl = 9$ for the bulge, consistent with that
expected for typical disc and bulge populations. The overall surface
brightness profile is shown by the solid line in Figure~\ref{f2}a,
with the disc and bulge contributions indicated by the dashed and
dot-dashed lines, respectively. The crosses are the radially averaged
measurements from Table~VI of Walterbos \& Kennicutt (1987). In
Figure~\ref{f2}b the solid, dashed and dot-dashed lines show the
overall, disc and bulge contributions to the rotation curve, with the
dotted line giving the halo contribution. The crosses are from Figure~2
of Kent (1989) and are based on the emission-line curves of Brinks \&
Shane (1984) and Roberts, Whitehurst \& Cram (1978). The fit to both
the surface brightness and rotation profiles is good, given the
simplicity of the models.

The lens and source velocities are described by rotational and random
components. The rotation velocity for each component is given in the
4th column of Table~\ref{t2}. The random motions are modeled by an
isotropic Gaussian distribution with a one-dimensional velocity
dispersion given by the 5th column. When calculating the relative
transverse lens speed $\vt$, we take account of both the motion of the
source and the observer. The observer is assumed to move in a circular
orbit about the centre of the Galaxy with a speed of 220~\kms. We do
not assume any relative transverse bulk motion between the Galaxy and
M31. In practice, only the observer's motion is of consequence for
Galaxy lenses, and only the source motion for M31 lenses.

Since one of the questions we wish to address is how well
pixel-lensing observables can characterize the MACHO mass, we shall
simply model the Galaxy and M31 MACHO mass distributions by a Dirac
$\delta$-function:
   \be
      \psi (\mh) \propto \frac{1}{\mh} \delta (m - \mh), \label{halomf}
   \ee
The stellar lens mass distribution in the disc and bulge is described
by a broken power law:
   \be
      \psi (\ms) \propto \left\{ \begin{array}{ll}
      \ms^{-0.75} & (\mlo < \ms < 0.5~\sm) \\
      \ms^{-2.2} & (0.5~\sm < \ms < \mup)
      \end{array}. \right. \label{starmf}
   \ee
The mass function is normalized to yield the same value for $\psi
(0.5~\sm)$ for either slope. We take a lower mass cut-off $\mlo =
0.08~\sm$ and an upper cut-off $\mup = 10~\sm$, corresponding closely
to the local Solar neighbourhood mass function \cite{gou97}. Whilst
this is a reasonable assumption for stars in the M31 disc, the mass
function will overestimate the contribution of massive stars in the
older bulge. The higher $\mtl$ assumed for the bulge also requires
that the disc and bulge mass functions be different. However, the
slope at high masses is steep, so the contribution of high mass stars
to the lensing rate is in any case small. Furthermore, as already
mentioned, we are only interested in the bulge population as a
contaminant of the halo lensing statistics. The choice of upper mass
cut-off for the bulge is therefore not critical for the present study,
so we simply adopt the same mass function for the disc and bulge.

The stellar components provide both lenses and sources. We assume that
the lens and source populations are the same and so described by the
same density, velocity and mass distributions.  For the disc and bulge
sources, we use the $V$-band luminosity function of Wielen, Jahreiss
\& Kr\"uger (1983) for stars with $M_V > 5$ and that of Bahcall \&
Soneira (1980) for $M_V \leq 5$. The two functions are normalized to
the same value at $M_V = 5$. A more detailed study of the M31
luminosity function is underway \cite{last00}.

\section{Predictions and trends for pixel lensing} \label{s5}

The simulations for the POINT-AGAPE survey are performed over 1, 3 and
10 observing seasons for 9 MACHO masses spanning the range $10^{-3} -
10~\sm$. Each simulation produces an estimate of the number of events
across the whole M31 disc for each lens component, together with a
library of typically $10^4$ candidates containing information such as
the lens position, duration and transverse velocity. Since $\te$
cannot generally be measured from the light-curve, we output both
$\te$ and $\tfw$. The event libraries can be filtered to provide an
estimate of the pixel-lensing rate for any field placement.

\subsection{Number of events} \label{s5.2}

\begin{figure}
\begin{center}
\epsfig{file=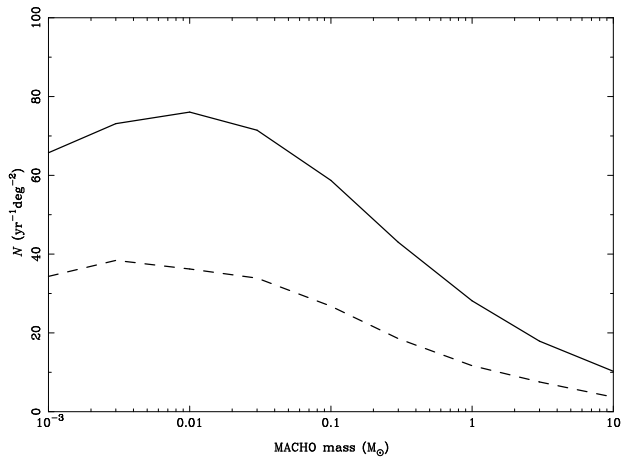,width=6cm,angle=270}
\end{center}
\caption{The expected event rate as a function of MACHO mass for full
MACHO haloes. The rates are averages over the M31 disc for M31 (solid
line) and Galaxy (dashed line) MACHOs and are computed from ten seasons
of data comprising 460 epochs.}
\label{nev}
\end{figure}
Whilst the factor $10^3$ gain over LMC/SMC searches in the number of
sources certainly boosts the rate of events, the fact that M31
pixel-lensing searches can typically detect only high-magnification
events means that the gain in the rate is not of the same
order. Nonetheless, as Figure~\ref{nev} indicates, the expected
pixel-lensing rate is almost an order of magnitude larger than for
current LMC/SMC experiments for same lens mass and halo fraction. In
the figure we have plotted the expected number of events for M31
MACHOs (solid line) and Galaxy MACHOs (dashed line) per season per
deg$^2$, assuming MACHOs comprise all the halo dark matter of both
galaxies. The rates are averages over the whole M31 disc (rather than
for a specific field placement) determined from simulations spanning
ten seasons and 460 observing epochs. Within the first season the
sensitivity to very massive MACHOs will be a little less than
indicated in Figure~\ref{nev}.

The rate of events occurring within the two INT WFC fields for their
first season (1999/2000) positions are displayed in
Table~\ref{t3}. This excludes events occurring within 5~arcmin of the
centre of M31 because this region is dominated by stellar self-lensing
(see Section~\ref{sdis}). Only a couple of self-lensing events per
season are expected outside the exclusion zone. The Monte-Carlo
error in the values in Table~\ref{t3} is small, only about $3\%$,
but one should expect a larger variation when comparing different
seasons with different numbers of epochs (in addition to Poisson
variations).

From Figure~\ref{nev} and Table~\ref{t3} we see that the sensitivity
to MACHOs peaks at a mass around $0.003-0.01~\sm$, when around 140
MACHO events can be expected within the INT WFC fields for full
haloes. Below $10^{-3}~\sm$ finite-source size effects become
important, so the expected number of events will drop off rapidly. At
the high mass end, even haloes comprising MACHOs as massive as
$10~\sm$ provide a rate of several events per season. The number of
M31 MACHOs is about twice as large as the number of Galaxy MACHOs for
the same mass and fractional contribution, which is a direct
consequence of the mass ratio of the halo models we adopt.

\begin{table}
\begin{center}
\caption{The expected number of M31 and Galaxy MACHO detections per
season (averaged over ten seasons comprising 460 epochs) for a range
of MACHO masses based on the placement of the two INT WFC fields in
the 1999/2000 observing season. The numbers assume the haloes of both
galaxies completely comprise MACHOs, though we exclude events occurring
within 5~arcmin of the centre of M31. For comparison, the expected
number of bulge and disc self-lensing events occurring outside the
exclusion zone is 2.2 per season. The Monte-Carlo error for a
given sequence of observing epochs is about $3\%$.}
\label{t3}
\begin{tabular}{@{}lcc}
Mass/$\sm$ & $N$(M31)/yr & $N$(Galaxy)/yr \\
\hline
0.001 & 87 & 38 \\
0.003 & 98 & 39 \\
0.01  & 97 & 37 \\
0.03  & 95 & 35 \\
0.1   & 76 & 28 \\
0.3   & 52 & 17 \\
1     & 32 & 12 \\
3     & 19 & 7.7 \\
10    & 10 & 3.1
\end{tabular}
\end{center}
\end{table}

\begin{figure}
\begin{center}
\epsfig{file=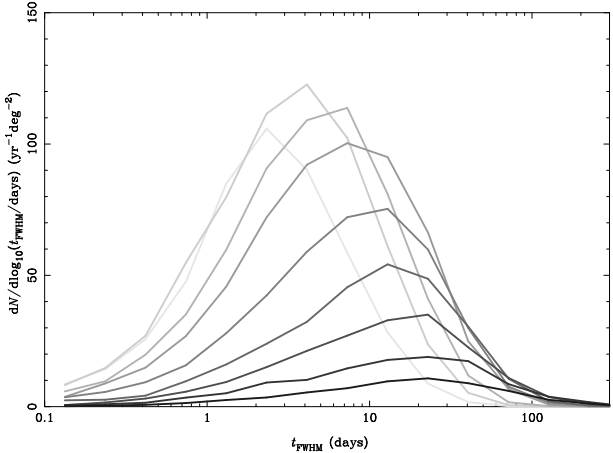,width=6cm,angle=270}
\end{center}
\caption{Observed MACHO timescale distributions for a range of MACHO
masses. The curves represent the combined M31 and Galaxy MACHO
normalized timescale distributions, shown in terms of the measured
FWHM timescale $\tfw$. From the lightest to the darkest curve the
MACHO mass is $0.001~\sm$, $0.003~\sm$, $0.01~\sm$, $0.03~\sm$,
$0.1~\sm$, $0.3~\sm$, $1~\sm$, $3~\sm$ and $10~\sm$.}
\label{t-dis}
\end{figure}

\subsection{Timescale distributions} \label{s5.2b}

In Figure~\ref{t-dis} we plot the timescale distributions for the
detected MACHOs for a range of masses in terms of $\tfw$.
The distributions for nine MACHO masses, spanning
four orders of magnitude, are plotted. The masses are as listed in
Table~\ref{t3}, with darker lines corresponding to more massive
MACHOs. Since the timescale distributions for Galaxy and M31 MACHOs
are practically indistinguishable for a given mass, in
Figure~\ref{t-dis} we have combined their timescale distributions, so
the normalization of each curve is determined by the combined
pixel-lensing rate shown in Figure~\ref{nev} for each halo.

\begin{table}
\begin{center}
\caption{Average timescales for M31 and Galaxy MACHO populations for a
range of masses. $\langle \tfw \rangle$ is the mean FWHM duration, as
measured from the light-curve, whereas $\langle \te \rangle_{\rm det}$
and $\langle \te \rangle_{\rm pop}$ are the mean Einstein radius
crossing durations of detected events and of the underlying
population, respectively.}
\label{t4}
\begin{tabular}{@{}lrcc}
Mass/$\sm$ & \null & M31 & Galaxy \\
\hline
0.001 & $\langle \tfw \rangle$: & 3.8 & 4.0 \\
\null & $\langle \te \rangle_{\rm det}$: & 6.2 & 6.3 \\
\null & $\langle \te \rangle_{\rm pop}$: & 2.3 & 3.1 \\
\null & \null & \null & \null \\
0.003 & $\langle \tfw \rangle$: & 5.1 & 5.1 \\
\null & $\langle \te \rangle_{\rm det}$: & 9.1 & 9.2 \\
\null & $\langle \te \rangle_{\rm pop}$: & 4.0 & 5.3 \\
\null & \null & \null & \null \\
0.01 & $\langle \tfw \rangle$: & 7.2 & 7.8 \\
\null & $\langle \te \rangle_{\rm det}$: & 14 & 15 \\
\null & $\langle \te \rangle_{\rm pop}$: & 7.3 & 9.7 \\
\null & \null & \null & \null \\
0.03 & $\langle \tfw \rangle$: & 9.7 & 9.4 \\
\null & $\langle \te \rangle_{\rm det}$: & 21 & 22 \\
\null & $\langle \te \rangle_{\rm pop}$: & 13 & 17 \\
\null & \null & \null & \null \\
0.1 & $\langle \tfw \rangle$: & 13 & 13 \\
\null & $\langle \te \rangle_{\rm det}$: & 34 & 37 \\
\null & $\langle \te \rangle_{\rm pop}$: & 23 & 31 \\
\null & \null & \null & \null \\
0.3 & $\langle \tfw \rangle$: & 16 & 17 \\
\null & $\langle \te \rangle_{\rm det}$: & 52 & 57 \\
\null & $\langle \te \rangle_{\rm pop}$: & 40 & 53 \\
\null & \null & \null & \null \\
1 & $\langle \tfw \rangle$: & 21 & 23 \\
\null & $\langle \te \rangle_{\rm det}$: & 82 & 98 \\
\null & $\langle \te \rangle_{\rm pop}$: & 73 & 97 \\
\null & \null & \null & \null \\
3 & $\langle \tfw \rangle$: & 26 & 28 \\
\null & $\langle \te \rangle_{\rm det}$: & 130 & 160 \\
\null & $\langle \te \rangle_{\rm pop}$: & 130 & 170 \\
\null & \null & \null & \null \\
10 & $\langle \tfw \rangle$: & 41 & 32 \\
\null & $\langle \te \rangle_{\rm det}$: & 220 & 300 \\
\null & $\langle \te \rangle_{\rm pop}$: & 230 & 310
\end{tabular}
\end{center}
\end{table}

\begin{figure}
\begin{center}
\epsfig{file=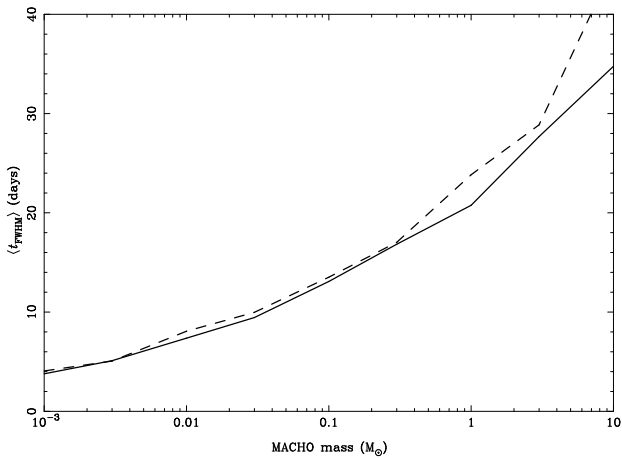,width=6cm,angle=270}
\end{center}
\caption{The mean FWHM duration, $\langle \tfw \rangle$, as a function
of MACHO mass. Line coding is as for Figure~\ref{nev}.}
\label{tfwav}
\end{figure}
\begin{figure}
\begin{center}
\epsfig{file=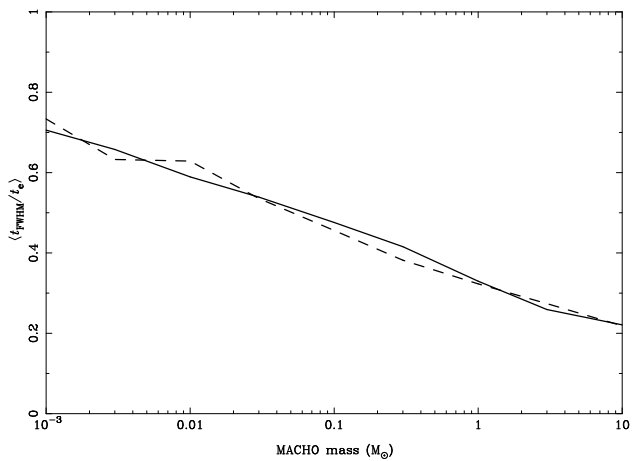,width=6cm,angle=270}
\end{center}
\caption{The mean ratio of FWHM duration to Einstein radius crossing
duration, $\langle \tfw / \te \rangle$, as a function of MACHO
mass. Line coding is as for Figure~\ref{nev}.}
\label{tfw-te}
\end{figure}
Whilst there is a clear trend of increasing $\tfw$ with increasing
MACHO mass, the correlation is much weaker than for $\te$. For example,
a duration $\tfw = 10-20$~days is typical of a $0.1~\sm$ lens, but it
is also not unusual for a lens as light as $10^{-3}~\sm$ or as heavy
as $10~\sm$.  Figure~\ref{tfwav} shows how the average duration
$\langle \tfw \rangle$ varies with mass separately for M31 (solid
line) and Galaxy (dashed line) MACHOs. Over four orders of magnitude
in mass $\langle \tfw \rangle$ varies by about one order of magnitude,
increasing from 4~days for $10^{-3}~\sm$ MACHOs to 35 days for
$10~\sm$ MACHOs (see also Table~\ref{t4}). For our sampling strategy
we find empirically that $\langle \tfw \rangle \propto m_{\rm
h}^{1/4}$, whereas the average Einstein radius crossing timescale for
the {\em underlying}\/ population of microlensing events (with $u
\leq 1$) scales as $\langle \te \rangle_{\rm pop} \propto m_{\rm
h}^{1/2}$.

The mean ratio $\langle \tfw / \te \rangle$ is displayed in
Figure~\ref{tfw-te} for detected events. It is clear that the ratio is
not fixed but steadily decreases with MACHO mass. For low MACHO masses
with short durations, sampling imposes a lower limit on $\tfw$ and a
loose lower limit on $\te$ as well. Whilst most events
involving $\sim 10^{-3}~\sm$ lenses are too short to be detected,
those that are either have an unusually long $\te$ or occur in
regions of low surface brightness (which maximizes $\tfw$ for a given
magnification). Thus $\langle \tfw / \te \rangle$ is typically larger
for the observed events. At the other end of the mass scale the
converse is true. The total observation baseline imposes a maximum
cutoff in $\tfw$ and a loose upper limit in $\te$. Those events which
are detected either have an unusually short $\te$ or else tend to
occur in regions of high surface brightness where $\tfw$ is minimized
for a given magnification. So $\langle \tfw / \te
\rangle$ tends to be smaller for observed events.  From
Table~\ref{t4} we see that the average duration of {\em detected}\/
events $\langle \te \rangle_{\rm det}$ does not trace the population
average $\langle \te \rangle_{\rm pop}$. This is a consequence of
sampling bias.

\subsection{Spatial distributions} \label{sdis}

\begin{figure*}
\centering
\begin{minipage}{170mm}
\begin{center}
\epsfig{file=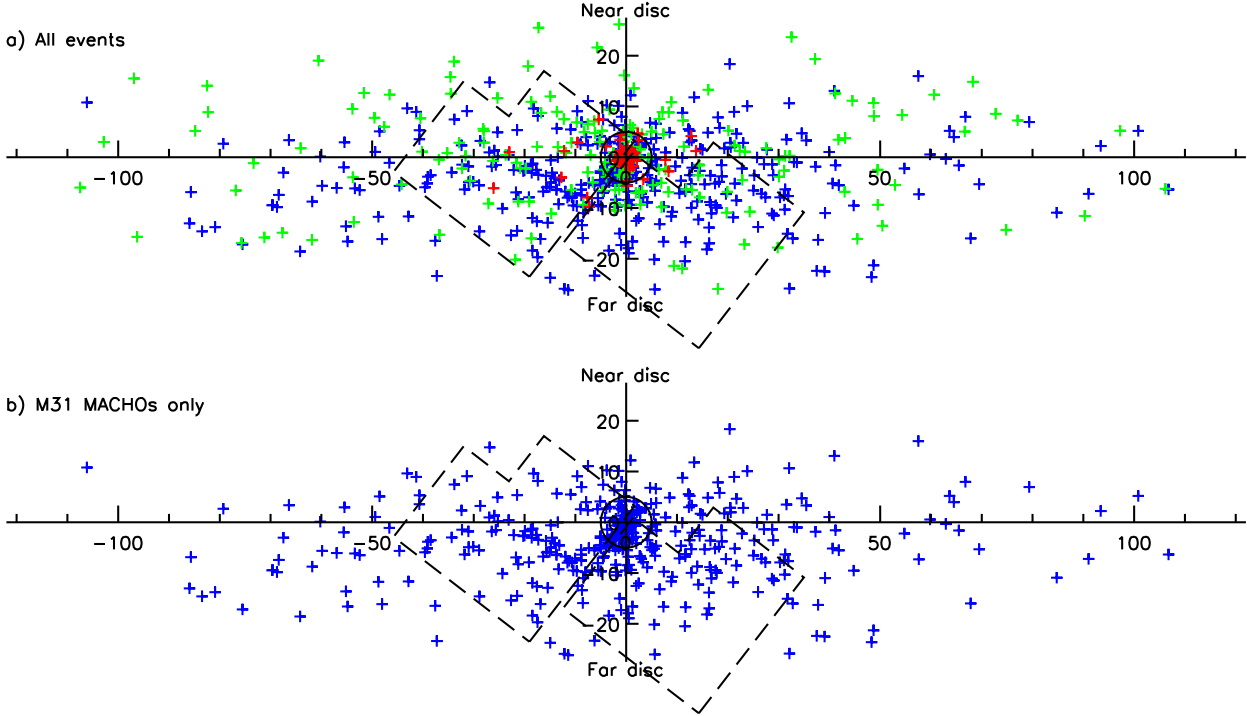,width=10cm,angle=90}
\end{center}
\caption{A realization for the spatial distribution of pixel-lensing
events after three seasons of observing, assuming MACHOs have a mass
of $0.3~\sm$ and provide all the halo dark matter in the Galaxy and
M31. The axis labelling is in arcmins. (a) The distribution of all
events. The green dots represent the foreground Galaxy MACHO
distribution, the red dots represent stellar lens events and
the blue dots depict the M31 MACHO distribution. The circle centred
on the origin demarcates the exclusion zone for the MACHO analysis,
inside which the rate is dominated by stellar self-lensing. The
dashed-line templates show the positions of the two INT fields for the
1999/2000 observing season. (b) The distribution of M31 MACHOs
only. The near-far asymmetry can be seen by comparing event number
densities at $\pm(10-20)$~arcmins along the minor axis.}
\label{f4}
\end{minipage}
\end{figure*}
\begin{figure*}
\centering
\begin{minipage}{170mm}
\begin{center}
\epsfig{file=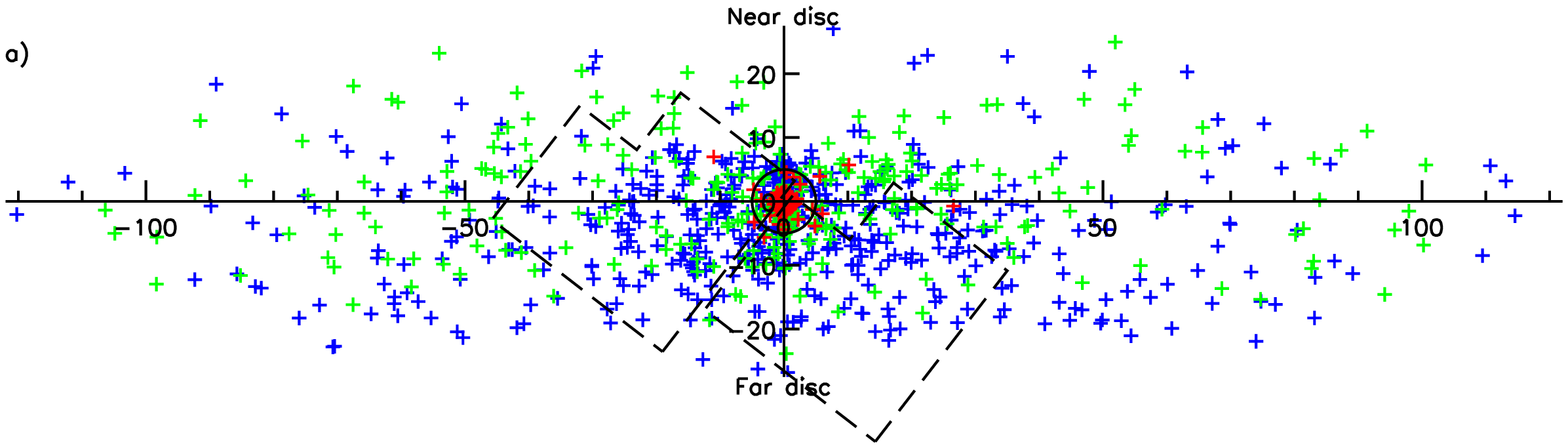,width=5cm,angle=90}
\epsfig{file=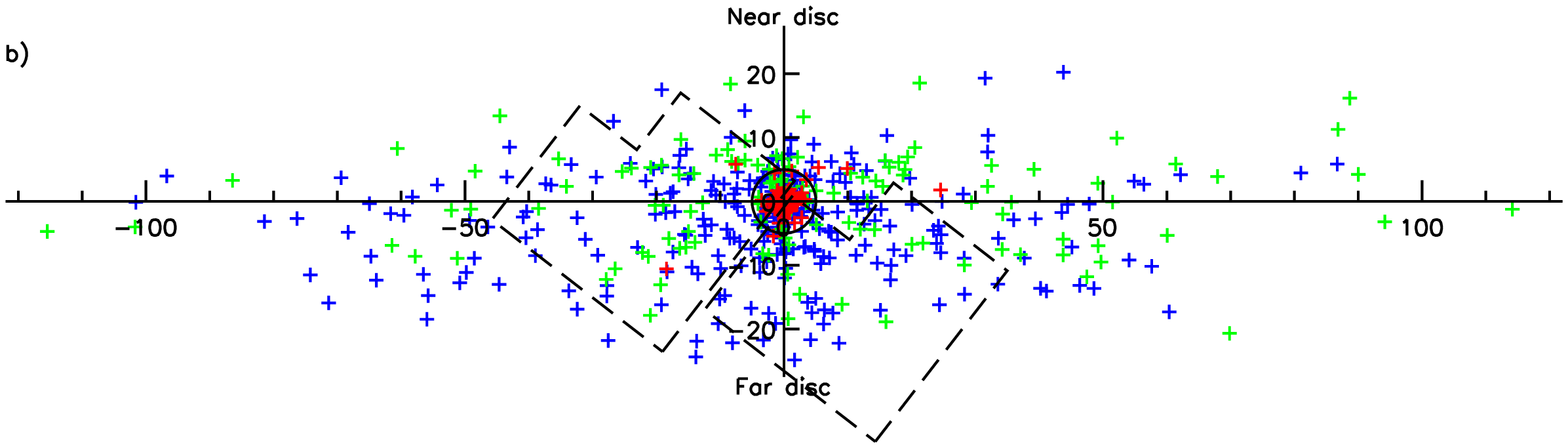,width=5cm,angle=90}
\epsfig{file=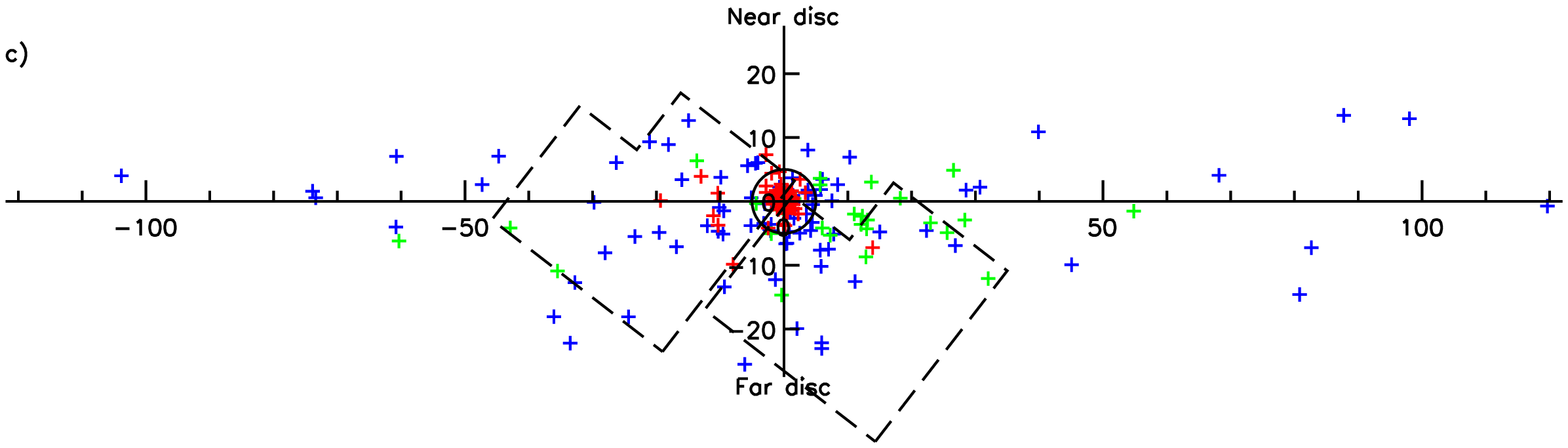,width=5cm,angle=90}
\end{center}
\caption{Realizations for the spatial distribution of pixel-lensing
events after three seasons of observing, assuming the MACHOs
in M31 and the Galaxy have the same mass and provide all the halo dark
matter. (a) The distribution for $0.1~\sm$ MACHOs; (b) $1~\sm$ MACHOs;
(c) $10~\sm$ MACHOs. The lines and symbols are as for Figure~\ref{f4}.}
\label{f5}
\end{minipage}
\end{figure*}
Since event timescales give only limited information in pixel lensing,
the location of each event on the sky is a crucial observable. A
robust measurement of near-far asymmetry in the event distribution
would indicate the existence of an extended spheroidal population of
lenses within which the visible M31 disc and bulge are embedded. Thus it
would represent very firm evidence for the existence of MACHOs.

In Figure~\ref{f4} we display the distribution of events across the
face of the M31 disc after three observing seasons for the case where
the haloes of both M31 and the Galaxy are full of $0.3~\sm$
MACHOs. The axes are labeled in arcmins and are aligned along the
major and minor axes of the disc light profile. The dashed-line
templates indicate the positions of the two INT WFC fields for the
1999/2000 observing season. 

In Figure~\ref{f4}a the positions of all detectable events are
shown. MACHOs from the Galaxy halo are shown in green whilst M31
MACHOs are shown in blue.  We find that within the central 5 arcmins
(denoted by the circle) most events are produced by ordinary stellar
lenses in the disc and bulge (shown in red). In Section~\ref{s6},
where we try to estimate MACHO parameters from simulated data-sets, we
disregard events occurring within this region so as to minimize
contamination from stellar lenses.

Figure~\ref{f4}b shows only the M31 MACHO distribution. The excess of
events between $y = -10$ and $-20$~arcmins (along the minor axis
towards the far side of the disc) compared to the number between $y =
+10$ and $+20$~arcmins is a consequence of near-far asymmetry in the
pixel-lensing rate. The strength of this asymmetry depends upon the
number of M31 MACHOs which, in turn, depends upon their mass and density
contribution, as well as the span of the
observation baseline. The presence of Galaxy MACHOs makes the
asymmetry harder to detect, so the ratio of M31 to Galaxy MACHOs is
another factor which determines whether or not the asymmetry is measurable.
It is evident from the figure that there are very few events at $|y| \ga
25$~arcmin. This is due to the decrease in both the number of sources
and the signal-to-noise ratio (because the sky background provides a
larger fraction of the total superpixel flux). The presence of the sky 
background effectively imposes a cut-off in the spatial distribution.

Figure~\ref{f5} shows the spatial distribution for a range of MACHO
masses expected after three seasons. We again assume that the MACHO
mass is the same in both galaxies and that MACHOs provide all the dark
matter in the two haloes. Figure~\ref{f5}a is for a MACHO mass of
$0.1~\sm$. In Figures~\ref{f5}b and \ref{f5}c the MACHO mass is
$1~\sm$ and $10~\sm$ respectively. The most obvious trend in the MACHO
distributions is the decrease in the number of detectable events for
models with more massive MACHOs. However, even for a mass as large as
$10~\sm$ we still expect to detect $30-40$ MACHOs within the INT
fields if they make up all the dark matter. After three seasons even
these massive MACHOs out-number the disc and bulge lenses lying
outside of our exclusion zone. This highlights one of the benefits of
pixel lensing: the reduction in $\tfw$ due to the presence of many
neighbouring unresolved sources means that more events with relatively
large $\te$ can be detected and characterized within a given observing
period. In this respect, pixel lensing is relatively more sensitive to
massive MACHOs than conventional microlensing experiments, which
require resolved sources.

Another noticeable trend in Figure~\ref{f5} is that more massive
MACHOs are concentrated towards the central regions of the M31
disc. The main reason is that the MACHO and source surface densities
are largest in this region, so the probability of an event occurring
there is larger. However, another factor is that it is in the regions
of highest surface brightness that the ratio $\tfw / \te$ is minimized
for a given magnification. For the $10~\sm$ MACHO model, where many
events may have a duration $\te$ exceeding the survey lifetime, this
means more light-curves can be fully characterized, enabling these
events to be flagged as microlensing candidates within the observing
period. The converse is true for low-mass MACHOs with short
$\te$. Their distribution is biased towards regions of lower surface
brightness where $\tfw / \te$ is maximized. This effect provides a
further degree of discrimination for different lens masses and means
that, for example, a halo with a modest contribution of low mass
MACHOs may be distinguished from one with a substantial fraction of
more massive lenses, even if the number of events for the two models
is comparable. This in part makes up for the fact that $\tfw$ is a
less powerful discriminant than $\te$.

\section{Estimating MACHO parameters} \label{s6}

In the previous section we found that, whilst the timescale
information in pixel-lensing studies is somewhat more restricted than
in conventional microlensing we do, at least for M31, have important
information from the spatial distribution of lenses. We now address to
what extent pixel-lensing observables permit a reconstruction of the
MACHO mass and halo fraction in the Galaxy and M31. 

\subsection{Maximum-likelihood estimation} \label{s6.1}

Alcock et al. (1996) presented a Bayesian maximum likelihood technique to
estimate the Galaxy MACHO mass and halo fraction from the observed
event timescales towards the LMC.  Evans \& Kerins (2000) extended this
to exploit the spatial distribution of observed events, and also to
allow for more than one significant lens population.  For pixel
lensing towards M31 we must also consider the effect of contamination
by variable stars. This is likely to be a significant problem in the
short term. A baseline of more than three years should be sufficient
to exclude periodic variables, such as Miras, but there still remains
the possibility that, occasionally, the signal-to-noise ratio may be
insufficient to distinguish between novae and microlensing events. By
taking account of variable stars in our likelihood estimator we
allow ourselves to make an estimate of the MACHO mass and lens
fraction which, even in the short term, is robust and unbiased.

In order to allow for different MACHO parameters in the two galaxies
we propose an estimator which is sensitive to five parameters: the
MACHO mass and halo fraction in both the Galaxy and M31, and
the degree of contamination by variable stars. We define our model
likelihood $L$ by
   \begin{eqnarray}
      \ln L(f_{\rm var},f_j,\psi_j) = & - & \left[ f_{\rm var} N_{\rm
      var} + \sum_{j=1}^{n_{\rm c}} f_j N(\psi_j) \right] \nonumber \\
      & + & \sum_{i = 1}^{N_{\rm obs}} \ln
      \left[ f_{\rm var} \frac{d^3 N_{\rm var}}{d{\tfw}_i dx_i dy_i}
      \right. \nonumber \\
      & + & \left. \sum_{j = 1}^{n_{\rm c}} f_j \frac{d^3
      N(\psi_j)}{d{\tfw}_i dx_i dy_i} \right], \label{like}
   \end{eqnarray}
where $f_{\rm var}$ is the fraction of variable stars relative to some
fiducial model expectation number $N_{\rm var}$, $f_j$ and $\psi_j$ are the
lens fraction and mass function for component $j$, $n_{\rm c}$ is
the number of lens components and $N_{\rm obs}$ the number of observed
events. For the disc and bulge components $f_j$ and $\psi_j$ are both
fixed, with $f_j = 1$ and $\psi_j$ given by equation~(\ref{starmf}),
whilst for the Galaxy and M31 haloes $\psi_j \propto m_j^{-1} \delta(m
- m_j)$, as in equation~(\ref{halomf}), and $f_j$ and $m_j$ are free
parameters. We define $f_j$ with respect to the halo density
normalizations in Table~\ref{t2}.

The resolution of our simulation is insufficient to evaluate reliably
the third derivatives in equation~(\ref{like}), so we decouple the
timescale and spatial distributions by computing $(dN/d\tfw) (d^2 N/dx
dy)$ instead of $d^3N/d\tfw dx dy$ within our fields. By averaging
over spatial variations in the timescale distribution we are ignoring
correlations which could provide us with further discriminatory
information. However, in the limit of infinite data and perfect
measurements we are still able to recover precisely the underlying
parameters because the average event duration is known with infinite
precision.

We assume that the distribution of variable stars traces the
M31 surface brightness. In reality variable stars will be harder to
detect in regions of higher surface brightness, so our idealized
distribution is somewhat more concentrated than we should expect for a
real experiment. We assume the timescale distribution of detectable
variables is log-normal, with a mean and dispersion $\langle \ln \tfw
\rangle = 2$ and $\sigma(\ln \tfw) = 0.5$ (where $\tfw$ is expressed
in days). Their timescales are therefore assumed to be typical of a
wide range of lens masses (see Figure~\ref{t-dis}) and are thus least
helpful as regards discrimination between lensing events and variable
stars.

To test the likelihood estimator we generate data-set realizations and
compute their likelihood over a five-dimensional grid of models
spanning a range of MACHO masses and variable star and MACHO
fractions. For the grid sampling we assume uniform priors in the
variable star and MACHO fractions and logarithmic priors for the MACHO
masses. Since the events in the inner 5~arcmin of the M31 disc are
predominately due to stellar lenses (mostly bulge self-lensing) we
count only events occurring outside of this region.

\subsection{First-season expectations}

\begin{figure*}
\centering
\begin{minipage}{170mm}
\begin{center}
\epsfig{file=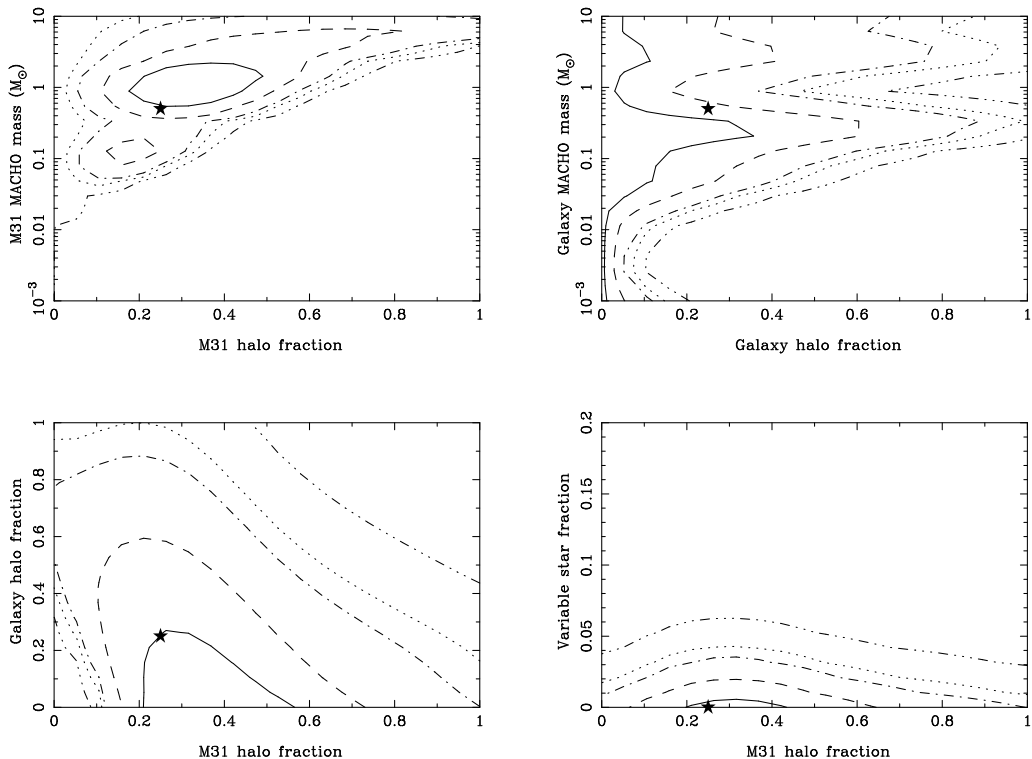,width=10cm,angle=270}
\end{center}
\caption{Maximum likelihood recovery of MACHO parameters for a
simulated data-set after one season. The input parameters for
both Galaxy and M31 MACHO populations are $0.5~\sm$ for the lens mass
and 0.25 for their fractional contribution. Here we assume there is no
contamination to the data-set from variable stars. The four panels are
two-dimensional projections of the five-dimensional likelihood
space. The contours in each plane enclose $34\%$ (solid line), $68\%$
(dashed line), $90\%$ (dot-dashed line), $95\%$ (dotted line) and
$99\%$ (triple dot-dashed line) of the total likelihood assuming a
linear prior in the MACHO and variable star fractions and a
logarithmic prior in the MACHO masses. The stars denote the input
parameters. The four panels represent the likelihood in the planes of
M31 MACHO fraction and mass ({\it top left}\/); Galaxy MACHO fraction
and mass ({\it top right}\/); M31 and Galaxy MACHO fractions ({\it
bottom left}\/); and M31 MACHO and variable star fractions ({\it bottom
right}\/). The variable star fraction is measured relative to a rate
of 100 events per year in the two INT fields.}
\label{f10}
\end{minipage}
\end{figure*}

Figure~\ref{f10} shows the degree to which the MACHO parameters can be
recovered after one season in the optimal case where the data-set
contains no variable stars. For the realization we have adopted a
MACHO fraction of 0.25 and mass of $0.5~\sm$ for both the Galaxy and
M31 haloes, and have set $f_{\rm var} = 0$. The MACHO parameters
correspond to those preferred by the most recent analyses of the EROS
and MACHO teams \cite{lass99,alc00}. Each panel in Figure~\ref{f10}
represents a two-dimensional projection of the five-dimensional
likelihood, in which each point on the two-dimensional plane is a
summation of likelihoods over the remaining three dimensions. Contours
are constructed about the two-dimensional maximum likelihood solution
which enclose a given fraction of the total likelihood over the
plane. The contours shown enclose $34\%$ (solid line), $68\%$ (dashed
line), $90\%$ (dot--dashed line), $95\%$ (dotted line) and $99\%$
(triple dot--dashed line) of the total likelihood. The star in each
plane shows the input values for the realization.

\begin{figure*}
\centering
\begin{minipage}{170mm}
\begin{center}
\epsfig{file=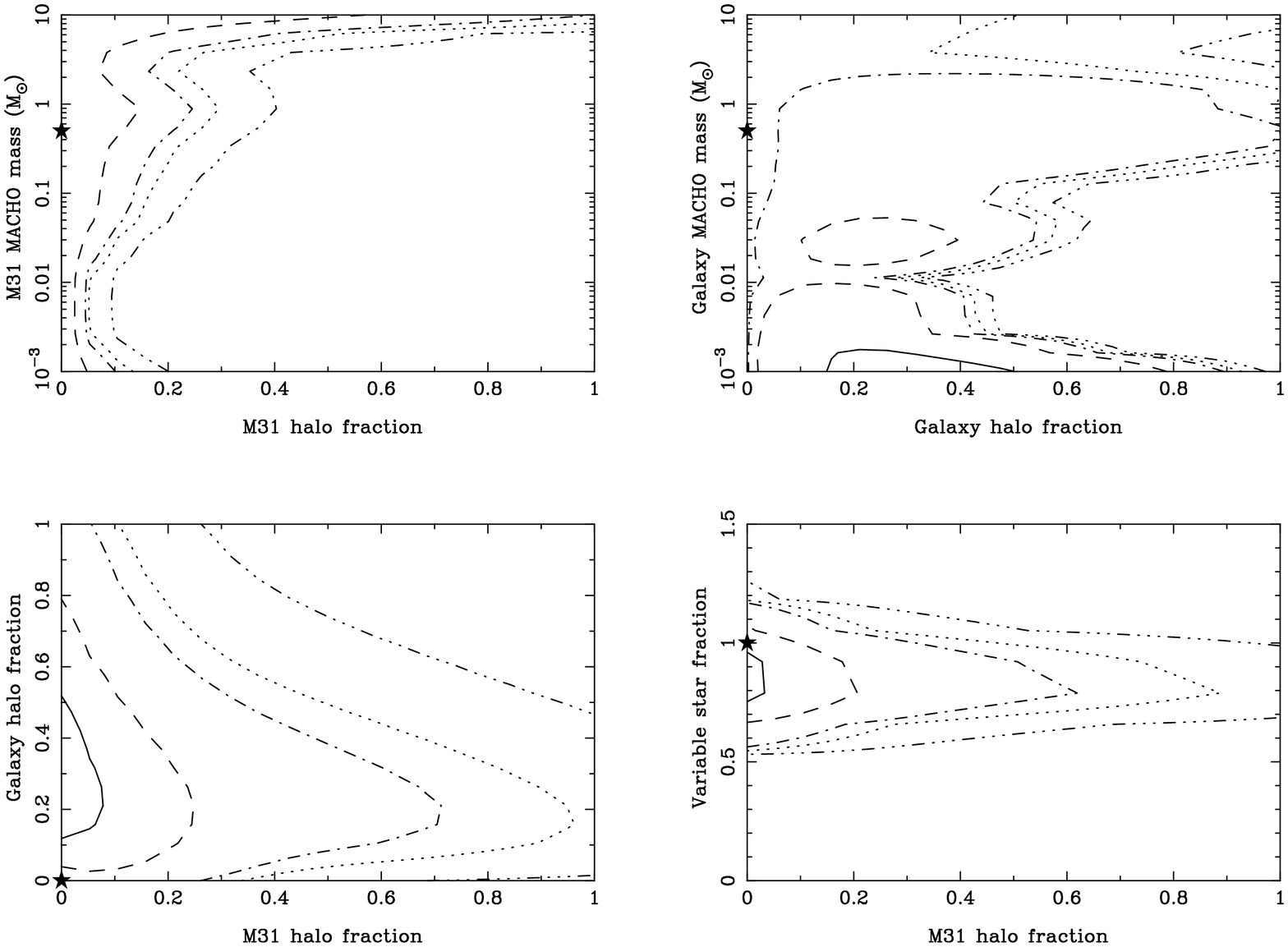,width=10cm,angle=270}
\end{center}
\caption{As for Figure~\ref{f10} but this time there are no
MACHOs, only variable stars. The input model has a variable star
fraction of unity.}
\label{f11}
\end{minipage}
\end{figure*}

\begin{figure*}
\centering
\begin{minipage}{170mm}
\begin{center}
\epsfig{file=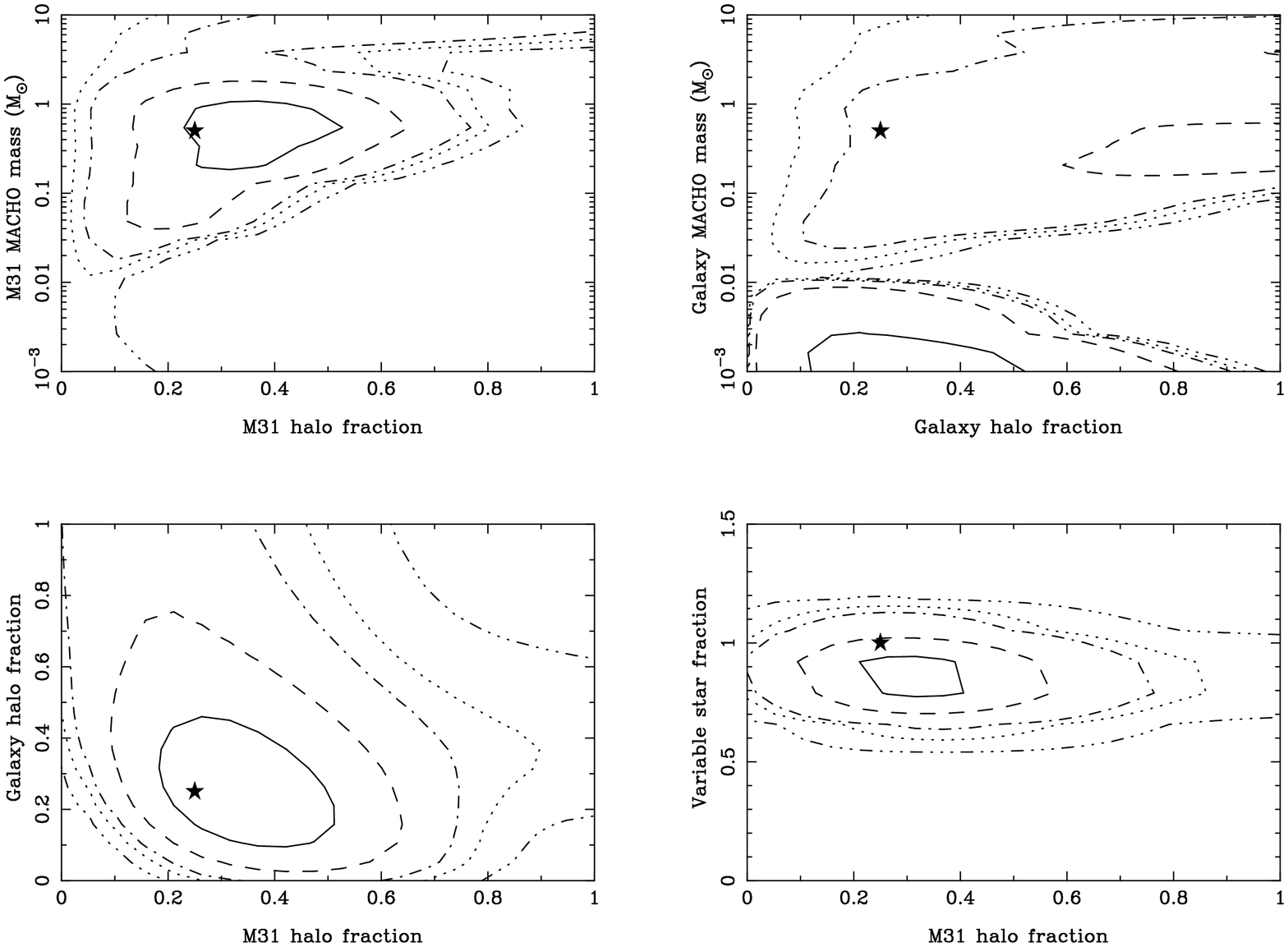,width=10cm,angle=270}
\end{center}
\caption{As for Figure~\ref{f10}, with the same input parameters, except
that we now adopt a variable star fraction of unity rather than
zero.}
\label{f12}
\end{minipage}
\end{figure*}

The four panels in Figure~\ref{f10} depict the likelihood planes for
M31 MACHO fraction and mass (top left), Galaxy MACHO fraction and mass
(top right), M31 and Galaxy MACHO fractions (bottom left) and M31
MACHO and variable star fractions (bottom right). From the top-left
panel we see that, after just one season, useful constraints are
already possible for M31 parameters. In this realization the $90\%$
confidence level spans around two orders of magnitude in MACHO mass
($\sim 0.05 - 10~\sm$) and an order of magnitude in halo fraction
($\sim 0.1 - 1.1$). The brown-dwarf regime is mostly excluded.  In the
upper-right panel we see that the Galaxy MACHO parameters are
ill-defined after one season. This is unsurprising since Galaxy MACHOs
are out-numbered two to one by M31 MACHOs and they have no signature
comparable to the near-far asymmetry of their M31 counterparts. The
panel shows a suggestive spike in the likelihood contours occurring at
about the right mass range, though the contours marginally prefer a
Galaxy halo with no MACHO component. The one firm conclusion that can
be drawn is that a substantial contribution of low-mass lenses is
strongly disfavoured by the data. The strongest constraints occur at
$\sim 0.003~\sm$, where the expected number of events peaks for a
given fractional contribution. The likelihood estimator indicates that
$0.003~\sm$ lenses contribute no more than $\sim 5\%$ of the Galactic
dark matter with $90\%$ confidence. In the lower-left and lower-right
panels of Figure~\ref{f10} we see the trade-off between M31 and Galaxy
MACHO fractions and between M31 MACHO and variable star fractions,
respectively. The lower-left panel indicates that a scenario in which
there are no MACHOs is excluded with very high confidence, despite the
large uncertainty in the halo fraction determinations. In the
lower-right panel we see that the likelihood estimator has correctly
determined that there is little, if any, contamination due to variable
stars, with a $90\%$ confidence upper limit of $f_{\rm var} < 0.03$.

In Figure~\ref{f11} we show the results for a simulation over one
season in which there are no microlensing events, only variable
stars. We adopt $N_{\rm var} = 100$ and $f_{\rm var} = 1$ within the
INT WFC fields. It is important to establish whether, in the event of
there being no MACHOs, our likelihood estimator is able to correctly
determine a null result even if a significant number of variable stars
pass the microlensing selection criteria. The four panels in
Figure~\ref{f11} indicate that our estimator has been very successful
as regards the M31 MACHO contribution. The M31 MACHO fraction is
constrained with $90\%$ confidence to be below 0.2 for lenses in the
mass range $0.001 - 0.1~\sm$ and below $0.4$ for MACHOs up to a few
Solar masses. This despite a rate in variable stars comparable to full
haloes of MACHOs. In the upper-right panel we see that there is
considerable uncertainty in the Galaxy MACHO parameters, though
interesting upper limits on the halo fraction are obtained for lenses
in the mass range $0.03 - 0.1~\sm$. In the lower-left panel we see
that a non-zero MACHO contribution is preferred though the contours
are consistent with the input model at about the $70\%$ confidence
level. In the lower-right panel we see that the estimator is able to
constrain the number of variables to within $\pm 30\%$ of the input
value. Thus our likelihood estimator has provided us with not just an
estimate of the MACHO parameters but also an estimate of the level of
contamination in the data-set. This estimate is completely independent
of (and thus does not rely upon) additional information one might
obtain from colour changes or asymmetry in the light-curves of
individual events, or from follow-up observations.

\subsection{Evolution of parameter estimation}

\begin{figure*}
\centering
\begin{minipage}{170mm}
\begin{center}
\epsfig{file=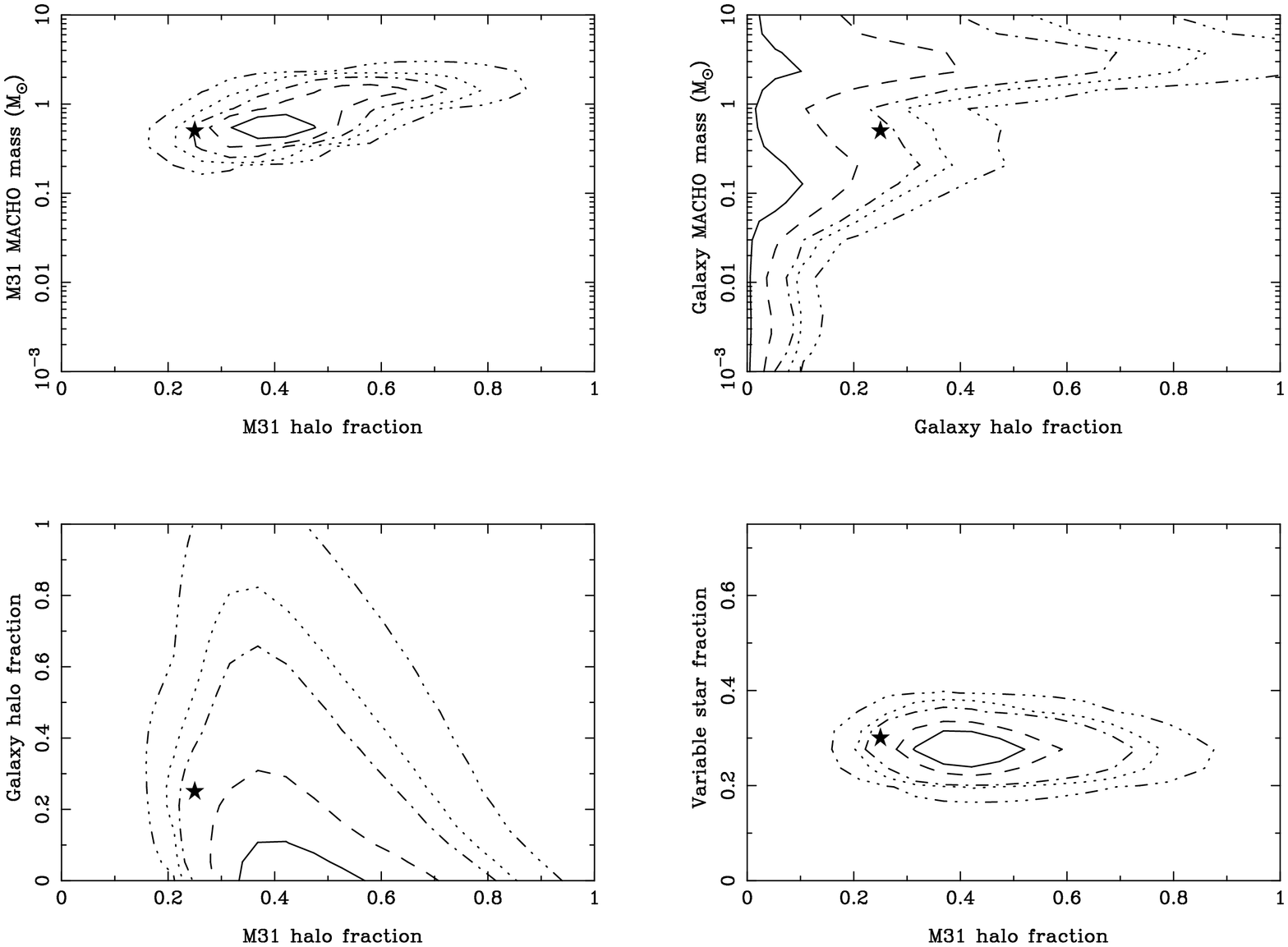,width=10cm,angle=270}
\end{center}
\caption{As for Figure~\ref{f12} but for three seasons of data and a
variable stars fraction of 0.3.}
\label{f13}
\end{minipage}
\end{figure*}

\begin{figure*}
\centering
\begin{minipage}{170mm}
\begin{center}
\epsfig{file=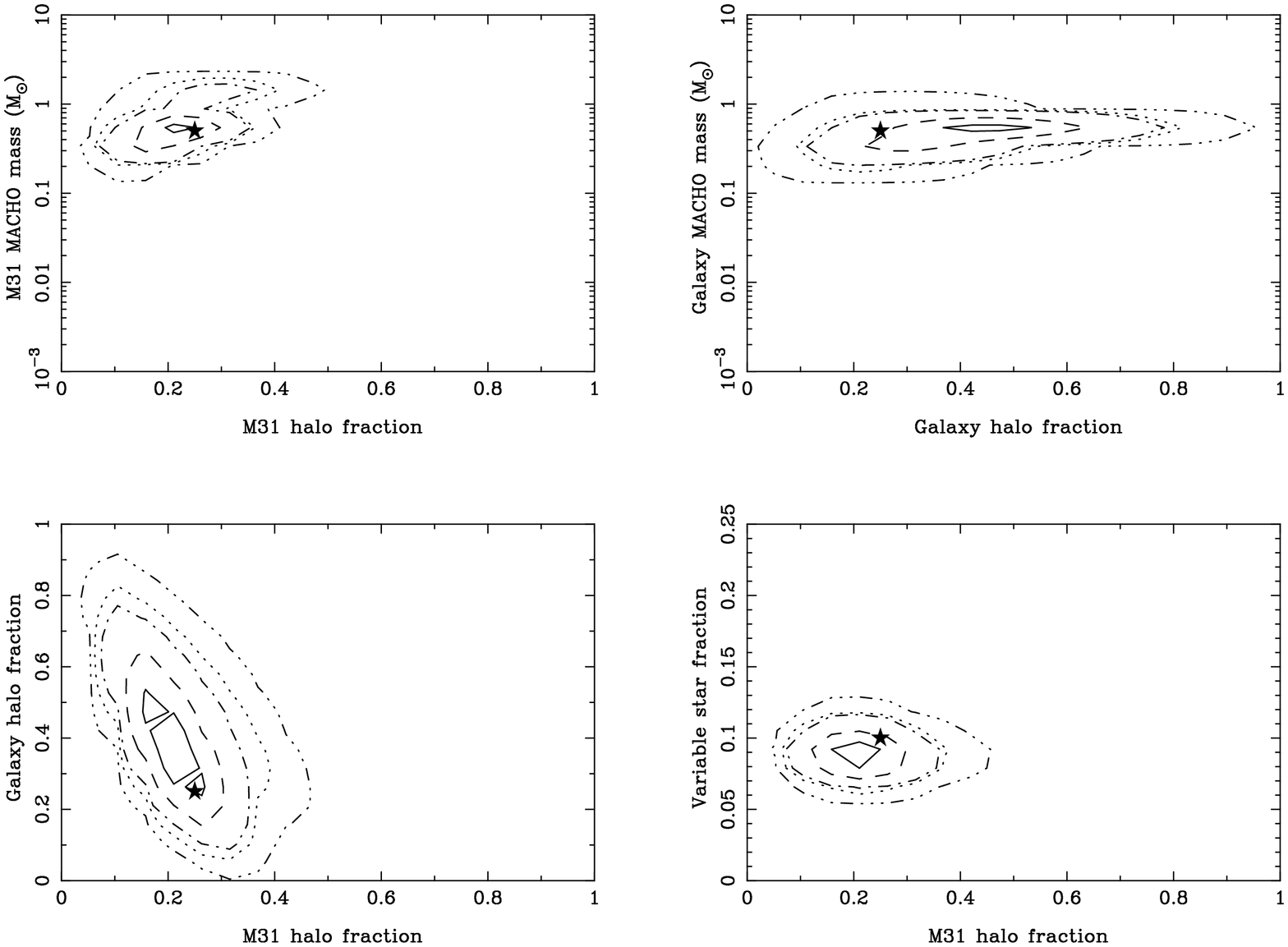,width=10cm,angle=270}
\end{center}
\caption{As for Figure~\ref{f12} but for ten seasons of data,
comparable to the lifetime of current LMC surveys, and a variable
stars fraction of 0.1.}
\label{f14}
\end{minipage}
\end{figure*}

Figure~\ref{f12} shows another first-season simulation in which we
adopt the same MACHO parameters as in Figure~\ref{f10} but this time
we also take $N_{\rm var} = 100$ and $f_{\rm var} = 1$. The contours
in the plane of M31 MACHO mass and fraction appear largely unaffected
by the presence of significant variable star contamination, and
qualitatively resemble those in Figure~\ref{f10}. There is no
evidence of estimator bias due to the presence of variables, which for
our realization out-number the MACHOs from both haloes
combined. However the Galaxy MACHO parameter estimation is clearly led
astray by the presence of variables, with upper limits on halo
fraction possible for only a narrow range of lens masses. The
estimator nonetheless strongly excludes a no-MACHO hypothesis
(lower-left panel) and provides a good estimate of variable star
contamination levels.

Figure~\ref{f13} shows the constraints after three seasons assuming
the same parameters as for Figure~\ref{f12}, except that we have
reduced the contamination level to $f_{\rm var} = 0.3$. A significant
decrease in contamination would be expected as the increase in
observation baseline permits the exclusion of a larger number of
periodic variables.  The constraints for M31 MACHO parameters have
tightened up considerably, with a $90\%$ confidence uncertainty of a
factor four in halo fraction and an order of magnitude in MACHO
mass. The constraints on Galaxy MACHO parameters have also sharpened
considerably, allowing strong upper limits on the halo fraction to be
made over a wide mass range, though the data in this case is
consistent with a complete absence of Galaxy MACHOs. However, in the
lower-left panel we see that the joint constraint on M31 and Galaxy
MACHO fraction advocates a significant overall MACHO contribution. The
lower-right panel also shows an accurate determination of
contamination levels.

In Figure~\ref{f14} we depict constraints for ten seasons of data,
comparable to the lifetime of current LMC surveys, with the variable
star contamination level reduced further to $f_{\rm var} = 0.1$. The
M31 MACHO fraction is now essentially specified to within about a
factor of three, whilst the MACHO mass uncertainty is
within an order of magnitude. We now also have a positive estimation
of the Galaxy MACHO contribution and mass. The constraints on Galaxy
parameters are only a little worse than those for M31 after three
seasons. The variable star contamination level is once again robustly
determined.

Figures~\ref{f12} to \ref{f14} show that the likelihood
estimator is able to distinguish clearly between microlensing events
and our naive model for the variable star population. They also show
that, given a lifetime comparable to the current LMC surveys, a
sustained campaign on the INT should determine M31 MACHO parameters
rather precisely and should also provide a useful estimate of Galaxy
MACHO parameters. A more modest campaign lasting three seasons would
provide a robust estimate of M31 MACHO parameters and useful
constraints on the Galaxy MACHO fraction.

Since all the above simulations assume the same halo fraction and
MACHO mass for both galaxies, we decided to test whether our
likelihood estimator was sensitive to Galaxy MACHO parameters
independently of M31 MACHO values. We therefore ran a simulation
over three seasons in which $30\%$ of the M31 halo comprises $0.5~\sm$
lenses and $60\%$ of the Galaxy halo comprises $0.03~\sm$ lenses. The
Galaxy MACHOs actually out-number the M31 MACHOs in this model. Whilst
the model is somewhat contrived, and is already ruled out with high
confidence \cite{lass99,alc00}, it provides a useful test case for
our estimator. We find that the estimator successfully resolves the
mass scales of the two populations within $90\%$ confidence, though
with a slight tendency to overestimate the Galaxy MACHO mass and
underestimate the M31 MACHO mass. Whilst we find a large overlap in
preferred halo fraction, this is consistent with the sensitivity
typically achieved after three seasons when the MACHO masses in the
two galaxies are the same. The Galaxy MACHO parameters are much better
defined than in Figure~\ref{f13}, though for this case the variable
star contamination level was set to zero.

There is one aspect, however, in which our simulation presents an
over-optimistic picture. The success of the estimator in
discriminating between variable stars and microlensing events is
mostly due to the fact that our adopted variable star distribution is
significantly more concentrated than the microlensing distribution of
either M31 or Galaxy MACHOs. The assumption we have made, that their
observed distribution traces the M31 surface brightness, is reasonable
only for very bright variable phenomena which would be detected
regardless of where it occurred in M31. For less prominent variables
there will be a bias against their detection in the central regions of
M31, where the surface brightness is high and so their contribution to
the superpixel flux relatively small. We might well expect a realistic
distribution of variable stars to resemble that of Galaxy MACHOs
because the surface density of Galaxy MACHOs does not vary
significantly over the M31 disc, so their distribution would also
trace the M31 surface brightness if there was no detection bias
away from regions of high surface brightness. However, in the absence
of a conspiracy between the flux distribution of variable stars at
peak luminosity and the flux distribution of microlensed sources at
peak magnification, there should be some distinction between the
spatial distributions of Galaxy MACHOs and M31 variable stars, though
this may be only mild. In any case, even if the two distributions are
indistinguishable this should not significantly affect the
determination of M31 MACHO parameters because the likelihood relies
heavily on evidence of near-far asymmetry (which is why the likelihood
contours are much better defined for M31 MACHOs than for Galaxy
MACHOs). This cannot be replicated by variable stars. Only if several
hundred variable stars passed the selection criteria every season would
the signature of asymmetry be washed out and the constraints on M31 MACHO
parameters severely degraded. Such an occurrence would warrant
critical re-examination of the selection criteria!

\section{Conclusions} \label{s7}

Pixel lensing is a relatively new and powerful method to allow
microlensing searches to be extended to targets where the sources are
unresolved. It heralds the possibility of detecting or
constraining MACHO populations in external galaxies. Though pixel
lensing is hampered by changes in observing conditions, which
introduce spurious variations in detected pixel flux, techniques have
been developed which minimize these variations to a level where genuine 
microlensing signals can be detected.

POINT-AGAPE and another team (MEGA) have embarked on a major joint
observing programme using the Isaac Newton Telescope (INT) to monitor
unresolved stars in M31 for evidence of pixel lensing due to MACHOs
either in the Galaxy or M31 itself. Two techniques, the Pixel Method
and difference imaging, are available to minimize flux variations
induced by the changing observing conditions. In this paper we have
assessed the extent to which the Pixel Method allows us to determine
the mass and fractional contribution of MACHOs in both M31 and the
Galaxy from pixel-lensing observables.  Our assessment takes account
of realistic variations in observing conditions, due to changes in
seeing and sky background, together with irregular sampling.

Pixel lensing observables differ from those in classical microlensing,
where one targets resolved stellar fields, in that one is generally
unable to measure the Einstein radius crossing time, $\te$, of an
event. The fact that the source stars are resolved only whilst they
are lensed means that one is unable to determine their baseline
luminosity, so neither the magnification nor the total duration of the
event can be measured. As an alternative to $\te$ one may measure the
full-width half-maximum timescale, $\tfw$, directly from the
light-curve. However, this provides only a lower limit to the
underlying event duration. Fortunately, M31 provides a signature which
permits an unambiguous determination of whether or not MACHOs reside
in its halo: near-far asymmetry \cite{cro92}. If M31 is embedded in a
dark spheroidal halo of MACHOs the high disc inclination should
provide a measurable gradient in the observed pixel lensing rate. The
strength of this signature depends both on the mass and fractional
contribution of MACHOs in M31, as well as the level of
``contamination'' by variable stars, M31 stellar lensing events and
foreground Galaxy MACHOs.

We have employed detailed Monte-Carlo simulations to
estimate the timescale and spatial distributions of MACHOs in both our 
Galaxy and M31 for spherically-symmetric near-isothermal
halo models. We also model the lensing contribution due to disc and 
bulge self-lensing. The expected number of M31 MACHOs for our two INT
fields peaks at about 100 events for $\sim 0.01~\sm$ MACHOs, the
Galaxy MACHO contribution being about half as large. For a given mass
and halo fraction we expect to detect about an order of magnitude more 
events than current conventional surveys targeting the LMC.

The timescale distributions for Galaxy and M31 MACHOs are practically
identical because of the symmetry of the microlensing geometry. Our
simulations also confirm that $\tfw$ is less strongly correlated with lens
mass than $\te$. For our sampling we find that, empirically, $\langle \tfw
\rangle \propto \langle \te \rangle^{1/2} \propto m^{1/4}$ for lens
mass $m$.  Sampling introduces a significant bias in the duration of
detected events with respect to the underlying average for very
massive and very light MACHOs.

Our simulations clearly show the near-far asymmetry in the M31 MACHO
spatial distribution. However, the presence of the foreground Galaxy
MACHOs makes its measurement more difficult. We also find that the
distribution of very massive MACHOs is noticeably more centrally
concentrated than that of less massive lenses.  Stellar self-lensing
events are found to be mostly confined to within the inner 5~arcmin of
the M31 disc, and are mostly due to bulge self-lensing. Their tight
spatial concentration means that they do not pose a serious
contamination problem for analysis of the Galaxy and M31 MACHO populations.

We have constructed a maximum likelihood estimator which uses
timescale and position observables to simultaneously constrain the
MACHO mass and halo fraction of both M31 and the Galaxy. The statistic
is devised to be robust to data-set contamination by variable
stars. We find that M31 MACHO parameters can be reliably constrained
by pixel lensing. For simulated INT data-sets we find pixel-lensing
constraints on the M31 halo to be comparable to those obtained for the
Galaxy halo by the conventional microlensing surveys. Even with severe
contamination from variable stars the M31 MACHO parameters are well
determined within three years. In particular, if there are few MACHOs
in M31 this should become apparent after just one season of data
collection, even if as many as a hundred variable stars pass the
microlensing selection criteria, because of the absence of near-far
asymmetry. Pixel lensing is less sensitive to Galaxy MACHO
parameters. Our simulations indicate that we require at least three
times as much observing time in order to produce comparable
constraints on Galaxy MACHO parameters. If the spatial distribution of
variable stars closely follows that of Galaxy MACHOs, then it may
become very difficult to reliably constrain Galaxy MACHO parameters.

The work presented here clearly demonstrates that a vigorous
monitoring campaign on a 2m class telescope with a wide-field camera
can identify and characterize MACHOs in M31. We now have the
opportunity to unambiguously establish the existence or absence of
MACHOs in an external galaxy.  The advantage of targeting M31 over our
own Galaxy is that we have many lines of sight through the halo of M31
and a clear signature with which to distinguish M31 MACHOs from
stellar self-lensing, the primary source of systematic
uncertainty for Galaxy halo microlensing surveys.  M31 therefore
represents one of the most promising lines of sight for MACHO studies.

\section*{acknowledgments}
EK, EL and SJS are supported by PPARC postdoctoral fellowships. NWE is
supported by the Royal Society. EK would like to thank Yannick
Giraud-H\'eraud and Jean Kaplan for many helpful discussions.

\end{document}